\documentclass[aps,superscriptaddress,twocolumn]{revtex4-1}
\usepackage{amsmath,amssymb,eucal,graphicx,color,units}

\begin{document}

\title{Spectral fingerprints of non-equilibrium dynamics: The case of a Brownian gyrator}

\author{Sara Cerasoli}
\affiliation{Department of Civil and Environmental Engineering, Princeton University, 
Princeton NJ 08544, USA}
\author{Sergio Ciliberto}
\affiliation{Laboratoire de Physique (UMR CNRS 567246), Ecole Normale Sup\'{e}rieure,  All\'{e}e d'Italie, 69364 Lyon, France}
\author{Enzo Marinari}
\affiliation{Dipartimento di Fisica, Sapienza Universit\`{a} di Roma, P.le A.
Moro 2, I-00185 Roma, Italy}
\affiliation{INFN, Sezione di Roma 1 and Nanotech-CNR, UOS di Roma, P.le A.
Moro 2, I-00185 Roma, Italy}
\author{Gleb Oshanin}
\affiliation{Sorbonne Universit\'e, CNRS, Laboratoire de Physique Th\'eorique de la Mati\`{e}re Condens\'ee (UMR CNRS 7600), 4 place Jussieu, 75252 Paris Cedex 05, France}
\email{gleb.oshanin@sorbonne-universite.fr}
\author{Luca Peliti}
\affiliation{Santa Marinella Research Institute, Santa Marinella, Italy}
\author{Lamberto Rondoni}
\affiliation{Dipartimento di Scienze Matematiche, Politecnico di Torino, Corso Duca degli Abruzzi 24, 10129 Torino, Italy}
\affiliation{INFN, Sezione di Torino, Via P. Giuria 1, 10125 Torino, Italy}

\date{\today}
\begin{abstract}
The same system can exhibit a completely different dynamical behavior 
when it evolves in equilibrium conditions or when it is driven out-of-equilibrium by, e.g.,  
connecting some of its components to heat baths kept at different temperatures. 
Here we concentrate on an analytically solvable and experimentally-relevant 
model of such a system -- the so-called Brownian gyrator -- a two-dimensional nanomachine 
that performs a systematic, on average, rotation around the origin under non-equilibrium conditions, while 
no net rotation takes place under equilibrium ones. On this example, 
we discuss a question whether it is possible to distinguish between two types of a behavior judging 
not upon the statistical properties of the trajectories of components, but rather 
upon their respective spectral densities. The latter 
are widely used to characterize diverse dynamical systems and
are routinely calculated from the data using standard built-in packages. 
From such a perspective,  we inquire whether the power spectral densities possess some "fingerprint" properties 
specific to the behavior in  
non-equilibrium. 
We show that indeed one can conclusively 
distinguish between equilibrium and non-equilibrium dynamics by analyzing the 
cross-correlations between the spectral densities of both components in the short frequency limit, or from 
the spectral densities of both components evaluated at zero frequency. Our analytical 
predictions,  corroborated by experimental and numerical results, open a new direction for the analysis 
of a non-equilibrium dynamics. 
 \end{abstract}
\maketitle
\section{Introduction}
Non-equilibrium phenomena and their time dependent and steady states 
have been one of the main research topics in the last decades. Numerous facets of such phenomena
have been scrutinized by experimental, numerical and theoretical analyses.  Seminal results have 
been obtained, as reviewed in a number of papers (see, e.g., Refs.~\cite{Ruelle,evans,Lam1,Lam2,Seki,Udo,Gal,CilibReview,pug,PelPig}).  
Main objects of investigation were the Fluctuation Relations, derived for
deterministic as well as for stochastic processes, for both transient and steady states, in a variety of guises, that have received numerical and experimental confirmations \cite{CilibReview}. These results have led, e.g., to the development of a linear response theory for non-equilibrium systems \cite{RuelleResp,ColRonVulp}, as well as to the analysis of the behavior of different non-equilibrium systems beyond the linear response regime
\cite{ESdissipth,Typicality,zia,baiesi,baiesi2,baiesi3,cividini,review,sarr}. The first applications have been in biophysical systems and in nanodevices and, more recently,  for colloidal particles captured in an optical trap  \cite{WangEvans,BustaLipRit,Gomez}.
Experimental applications to \textit{macroscopic} systems are less developed but nonetheless there exist few examples for which such an analysis was successful~\cite{AURIGA,Breaking}.

Concurrently, a long-standing question is whether it is possible to extract some reliable information about a complex system operating at non-equilibrium conditions directly from experimentally observed quantities. 
Various approaches have been developed to detect a violation of the detailed balance, which is itself indicative of non-equilibrium dynamics (see e.g. \cite{gonz}), or to determine the rate of entropy production (see, e.g. \cite{harada,lander,roldan}), which provides an information of how far the system is from equilibrium. Apart from that, 
more sophisticated approaches have been proposed to quantify, in particular settings, the average phase-space velocity field \cite{frishman}, the thermodynamic force field \cite{esposito}  or the patterns of microscopic forces \cite{volpe}. A review of the activities in these directions can be found in recent Ref. \cite{manik}, which also employed the thermodynamic uncertainty relation to quantify the thermodynamic force field 
as well as the rate of entropy production from short-time trajectory data.

In this paper we address the question whether one can distinguish between dynamics under equilibrium or non-equilibrium conditions from a completely different perspective. Namely, we inquire here if it is possible to find an unequivocal criterion discriminating between these two situations  
 judging 
not upon the statistical properties of the trajectories of the system's degrees of freedom, but rather 
upon their respective spectral densities.   
To our knowledge, this question has not been addressed before.
We note that, in general, an analysis based on the ensemble-averaged 
power spectral densities is rather standard and there exist built-in packages that automatically create the latter from the experimentally-recorded sets of data. Such an analysis 
has been proven to provide a deep insight into the dynamical behavior  
and has 
 been successfully applied across  several disciplines, from quantum physics to cosmology, and from crystallography to neuroscience, and to speech recognition (see, e.g., Ref.~\cite{catch} and references therein). This kind of approach has been also used for the analysis of climate data~\cite{talk}, musical recordings~\cite{geisel}, blinking quantum dots~\cite{n1,n2}, 
as well as for the trajectories of anomalous diffusion~\cite{ped,g1,g2,vit1,vit2,RSI,enzo,alessio,alessio1,alessio2} or also some non-equilibrium systems \cite{harada,fodor,jap}. Here, however, we go beyond the standard definitions focusing on power spectral densities of individual trajectories  of the components, which are random variables, parameterized by the observation time $t$ and frequency $\omega$, aiming to determine their joint bivariate and marginal probability density functions and cross-correlation. 
One interesting consequence of our analysis is that these quantities allow us to know whether the system evolves under 
non-equilibrium conditions or not.

\begin{figure}[htbp]
	\begin{center}
		\includegraphics[width=0.48\textwidth]{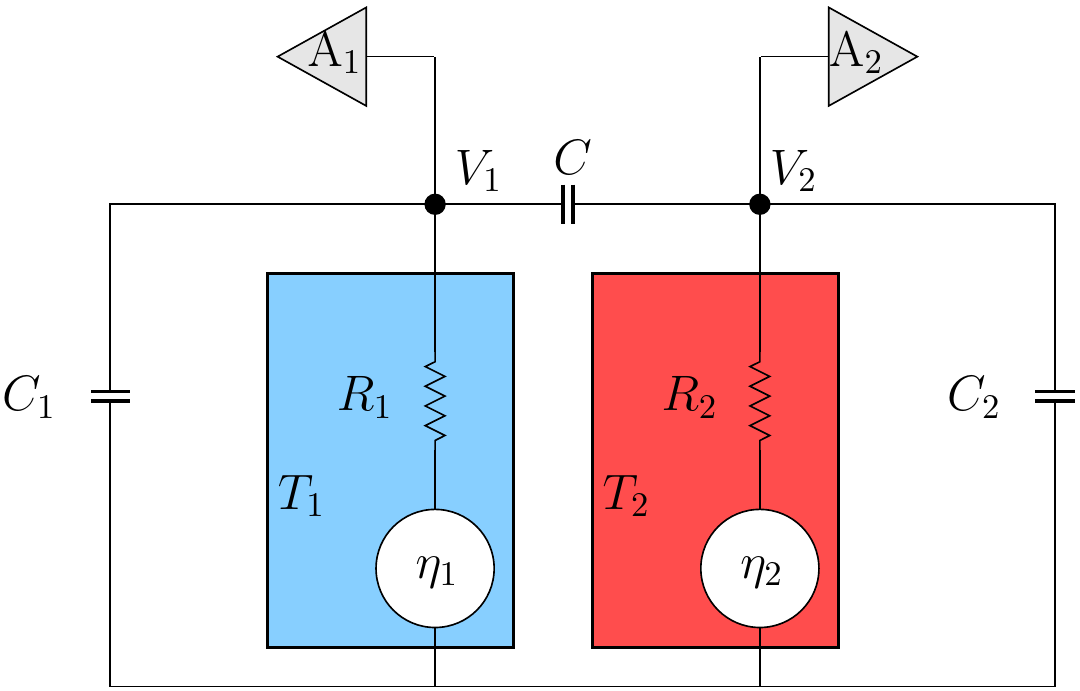}
	\end{center}
	\caption{(Color online) 
	A sketch of the circuit in Refs.~\cite{al1,al2}. The two resistances $R_1$ and $R_2$ are kept in separate screened boxes at the temperatures $T_1$ and $T_2$ respectively, and are coupled only by the capacitance~$C$. The signal consists of the voltages $V_1$ and $V_2$, which are measured via the amplifiers A$_1$ and~A$_2$, respectively. The noise sources $\eta_i$ and the capacitances~$C_i$, $i=1,2$, of the two circuits are also sketched.  The corresponding mathematical model is defined in Eq. \eqref{langevin_currents}.}
	\label{fig1}
\end{figure}

Here, we focus on the so-called Brownian gyrator modell \cite{pel,filliger}, which represents  a minimal two-dimensional nano-machine which on average experiences
persistent rotations, when kept in non-equilibrium conditions, while no rotation takes place when the system evolves under equilibrium conditions. 
This model is sufficiently simple to be analytically solved  and to obtain all the properties of interest in an explicit form,
showing at the same time a non-trivial behavior. Therefore, our investigation provides some insight on the behavior in
more complex situations, not directly included in our analysis. In general, for such complex sistuations only a numerical analysis is possible,
not only of the power spectral densities considered here, but also of other properties, such as work done by forces, entropy production and etc.
Moreover, this model can be realized experimentally, allowing our analytical predictions and concepts to be tested against experimental data.
To this end, we focus here on a particular version of the Brownian gyrator model as experimentally investigated in 
Refs.~\cite{al1,al2} (see Fig.~\ref{fig1}). This model consists of two electrical resistances kept at different temperatures and coupled 
only by the electrical thermal fluctuations, i.e. the Nyquist noise, which generates a heat flux between the two heat baths. 
The statistical analysis of the power spectral density of individual realizations of the measured noisy voltages allows us to 
fully discriminate between equilibrium dynamics and the non-equilibrium one. 
We also remark that our analysis 
can be easily adapted to other experimental realizations of the Brownian gyrator model, such as, e.g., a single colloidal particles suspended in an aqueous solution and confined by potentials generated by optical tweezers \cite{11}, by substituting the corresponding values of the parameters in the evolution equations.

The paper is organized as follows: In Sec.~\ref{model}, we define our model and notations in the context of the experimental realization of the Brownian gyrator in Refs.~\cite{al1,al2,CilibReview}. We also summarize the main results of our analysis. In Sec.~\ref{main}, we define the observables that allow us to discriminate equilibrium from non-equilibrium. In Sec.~\ref{dist} we evaluate the exact expressions for the bivariate moment-generating and for the bivariate probability density functions 
of single-trajectory power spectral densities.  In Sec.~\ref{conc} we summarize our results and discuss their physical significance. The details of the calculations are presented in the Appendices.

\section{The Brownian Gyrator}
\label{model}
\subsection{The experimental set-up}
The experimental set-up of Refs.~\cite{al1,al2} is sketched in Fig.~\ref{fig1}. It is made of two resistances $R_1$ and $R_2$, which are kept at different temperature $T_1$ and $T_2$, respectively. These temperatures are controlled  by thermal baths and $T_2$ is kept fixed at $296\,\unit{K}$ whereas $T_1$  can be set at any value  between $296\,\unit{K}$ and $88\,\unit{K}$, using the stratified vapor above a liquid nitrogen bath. 
In the figure, the two resistances have been drawn with their associated thermal noise generators $\eta_1$ and $\eta_2$, whose power spectral densities are given by the Nyquist formula $\overline{|\tilde \eta_i|^2}= 2 k_\mathrm{B} T_i\,R_i$, with $i=1,2$.    The coupling capacitance $C$ controls the electrical power exchanged between the resistances  and as a consequence the energy exchanged between the two baths. No other coupling exists between the two resistances which are inside two separated screened boxes.
The quantities $C_1$ and $C_2$ are the capacitances of the circuits and the cables.
Two extremely low noise amplifiers  A$_1$ and A$_2$ \cite{RSI} measure the voltage $V_1$ and $V_2$ across  the resistances $R_1$ and $R_2$ respectively. All the relevant quantities considered in this paper can be derived by the measurements of $V_1$ and $V_2$, as discussed below. 
 Mathematically, such a system is described by a pair of coupled Ornstein-Uhlenbeck processes at respective temperatures $T_1$ and $T_2$. When $T_1 \neq T_2$ the system evolves in non-equilibrium conditions and eventually
 reaches a non-equilibrium steady state. If $T_1=T_2$, the system  reaches equilibrium.  
 
 \begin{figure}[htbp]
	\begin{center}
		\includegraphics[width=0.46\textwidth]{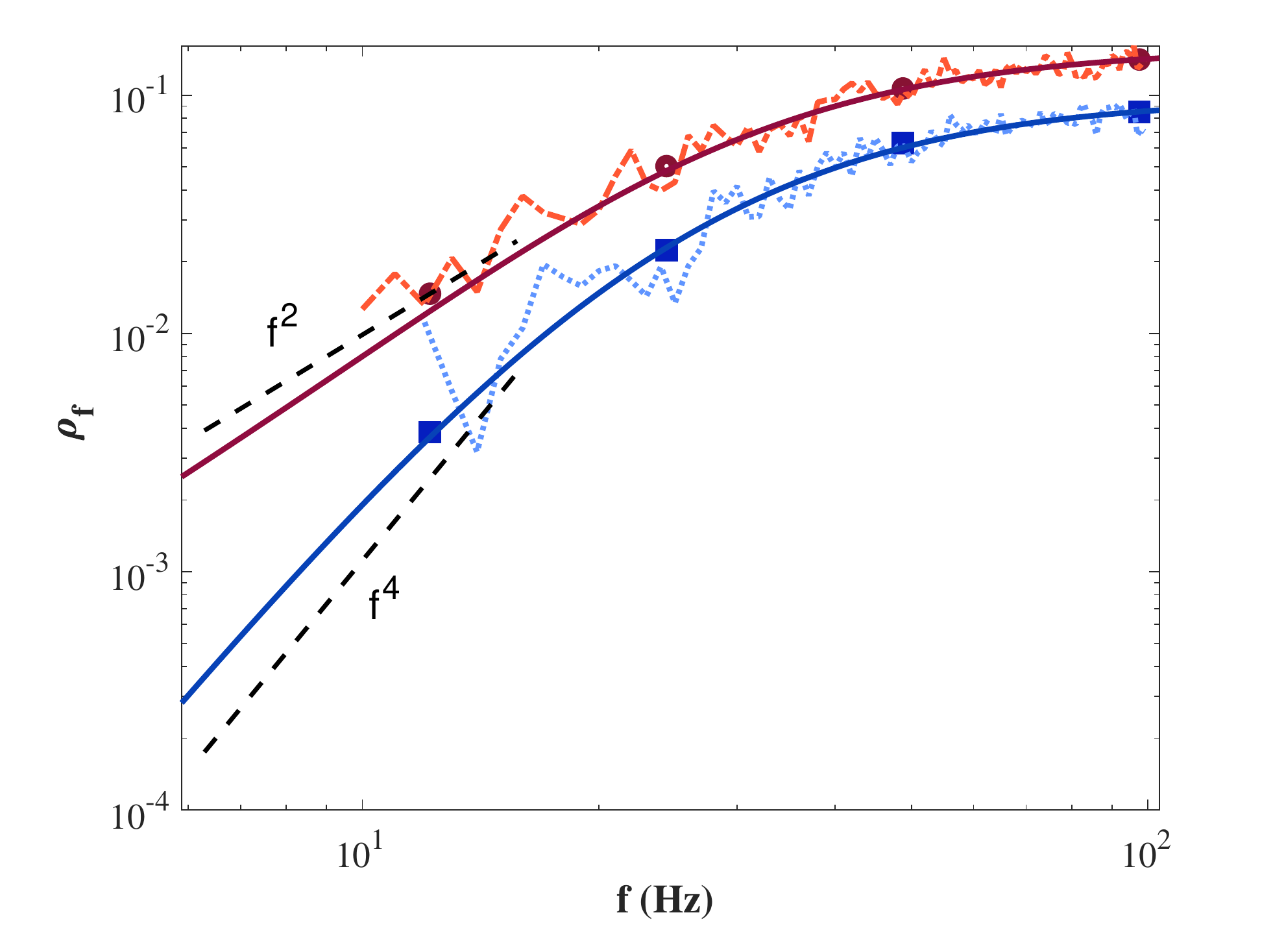}
			\end{center}
	\caption{(Color online) The Pearson coefficient $\rho_\omega$ in Eq. \eqref{pearson2} (solid curves) as a function of the frequency $f = \omega/2 \pi$. 
	The noisy curves present $\rho_{\omega}$ evaluated from the experimental data corresponding to the  following values of the system 
	parameters: $C = 100 \, \unit{pF}$, $C_1 = 680 \, \unit{pF}$ and $C_2 = 430 \, \unit{pF}$, and $R_1 = R_2 = 10\,\unit{M\Omega}$. For non-equilibrium dynamics 
we have $T_1 = 88 \, \unit{K}$ and  $T_2 = 296 \, \unit{K}$, while for equilibrium dynamics the system was maintained at equal temperatures  $T_1 = T_2 = 296 \, {\rm K}$.
	Solid line: theoretical prediction, Eq.~\eqref{pearson2}. Symbols: numerical simulations, average over 300.000 histories.  Dotted lines: experiment,  resolution $1 \, \unit{Hz}$.  Black dashed: frequency dependence for small $f$.} 
	\label{fig3}
\end{figure}

\begin{figure}[htbp]
	\begin{center}
	 \includegraphics[width=0.47\textwidth]{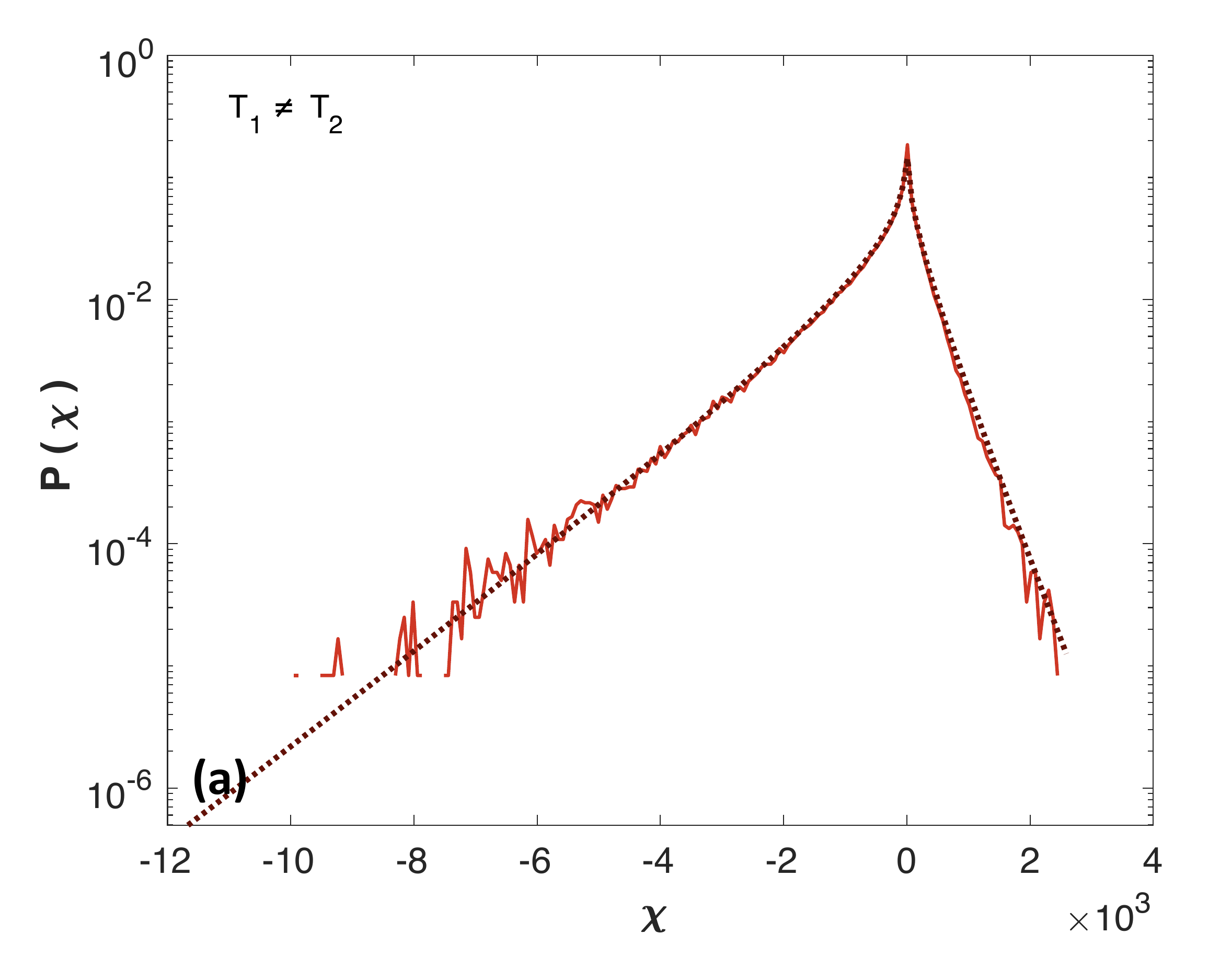}
		\includegraphics[width=0.47\textwidth]{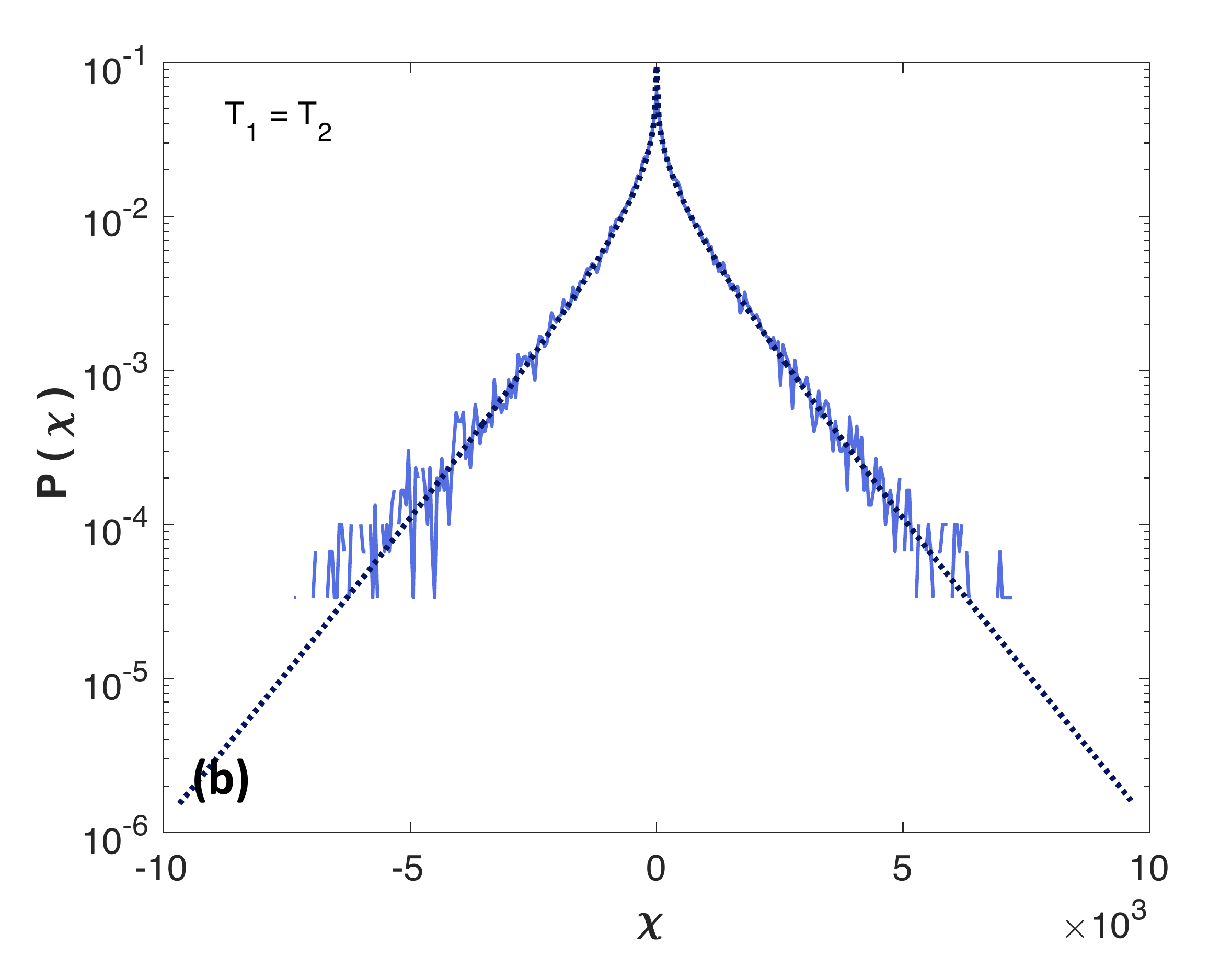}
			\end{center}
	\caption{(Color online) Probability density function (see Eq. \eqref{distchi}) of the random variable $\chi$ (see Eqs. \eqref{chi} and \eqref{Rparameter} for the definition) for non-equilibrium (panel (a)) and equilibrium (panel (b)) situations. Theoretical prediction \eqref{distchi} (solid thick line), thin line (experiment). Circuit parameters and values of the temperatures for equilibrium and non-equilibrium cases are the same as in Fig. \ref{fig3}.}
	\label{fig5}
\end{figure}

\subsection{Main analytical and experimental results}

Before going into all the mathematical analyses we want first to show that 
 it is indeed possible to discriminate the
	equilibrium ($T_1 = T_2$) from non-equilibrium ($T_1 \neq T_2$) dynamics from the measurement of realization-dependent spectral densities of the components of individual trajectories. In order to do that we analyze the individual realizations of the measured voltages $V_1$ and $V_2$,  
	as functions of the time $\tau$ within the observation interval $(0,t)$, and evaluate the corresponding power spectral densities $S_1(\omega,t)$ and $S_2(\omega,t)$, with $\omega=2\pi f$. The first possibility of discriminating between in and out of equilibrium  
	is provided by the 
	cross-correlations of the realization-dependent 
	spectral densities of both components (a quantity not usually considered), which permits to determine the Pearson correlation coefficient $\rho_{\omega}$ (see the definition in Eq. \eqref{pearson2} in the text). As evidenced in Fig. \ref{fig3}, the latter 
	exhibits  a strikingly different behavior at small frequencies in equilibrium and in non-equilibrium situations; Namely, in equilibrium $\rho_{\omega} \sim \omega^{4}$, while in non-equilibrium we have a slower power-law dependence $\rho_{\omega} \sim \omega^{2}$. 
	The second possibility is provided by the characteristic parameter $\chi$, which is a random variable defined as a suitably rescaled difference of the single-trajectory power spectral densities of the components evaluated at zero frequency (for the precise definition see Eqs. \eqref{chi} and \eqref{Rparameter}). We show that, on average, $\chi$ is identically equal to zero in equilibrium, and conversely, has a non-zero value in non-equilibrium. The full probability density function of this parameter is presented in 
	Fig. \ref{fig5} and it shows a strong asymmetry of the tails in the non-equilibrium state. This asymmetry disappears in equilibrium.  Besides these two  main results depicted in Figs. \ref{fig3} and \ref{fig5}, we also determine
	the explicit expressions of the full bivariate and univariate probability density functions of single-trajectory realization-dependent spectra, which lead to an interesting characterizations of the spectral densities of individual trajectories. We will show in particular that for an infinite observation time a single-trajectory spectral density at a given frequency is given by the first moment of this random function multiplied by a dimensionless random number whose distribution is frequency-independent and is explicitly evaluated.  This extends to localized Ornstein-Uhlenbeck-type processes, such that the mean-squared value of the random variable under study approaches a constant value as time tends to infinity, the earlier results for random processes for which such mean-squared value diverges with time~\cite{g1,g2,vit1,vit2}.
	In the next sections we will describe how these theoretical results and those plotted in Figs.\ref{fig3} and \ref{fig5}  have been obtained.

\subsection{The model}

In the circuit of Fig.\ref{fig1}, the total charges $q_i(\tau)$  that pass through the resistance $R_i$, $i=1,2$, in a time interval of duration $t$, satisfy the Langevin equations
\begin{equation}
\label{langevin_currents}
\begin{split}
R_1 \, \dot{q}_1 &= - \frac{C_2}{X} q_1 + \frac{C}{X} \left(q_2 - q_1\right) + \eta_1 \,, \\
R_2 \, \dot{q}_2  &= - \frac{C_1}{X} q_2 + \frac{C}{X} \left(q_1 - q_2\right) + \eta_2  \,, 
\end{split}
\end{equation}
with the initial conditions $q_1(0) = q_2(0) = 0$. In Eqs. \eqref{langevin_currents} we have defined 
$X = C C_1 + C C_2 + C_1 C_2$, while the noises $\eta_i$ have zero mean and covariance functions defined by
\begin{align}
\label{noises}
\overline{\eta_i(t) \eta_j(t')} = 2 k_\mathrm{B} T_j R_j\, \delta_{i,j}\, \delta\left(t-t'\right)\,,\qquad  i,j = 1,2 \,,
\end{align}
where $k_\mathrm{B}$ is the Boltzmann constant, and the overbar, here and henceforth, denotes the average over the realizations of noises. 

This model allows for different physical interpretations. Indeed, $\mathbf{q}=(q_1,q_2)$ can be regarded as the position of a particle undergoing Langevin dynamics on a plane, in the presence of a parabolic potential and of Gaussian noises with different amplitudes along the $q_1$- and $q_2$-directions. 
Such a system has been first analyzed in Ref.~\cite{pel}. It was later observed~\cite{filliger} that it represents a minimal model of a heat machine: whenever $T_1 \neq T_2$, a systematic torque is generated, and the particle acts as a ``Brownian'' gyrator, that experiences persistent (on average) rotations around the origin,  whose direction is defined by the sign of the difference $T_1 - T_2$ of the temperatures. For $T_1 = T_2$, the thermodynamic equilibrium correctly does not allow any net rotations.

This observation motivated extensive investigations of the BG model, see, e.g.,
Refs.~\cite{al1,al2,crisanti,2,12,11,3,alberto,bae,13,Lam3,Lam4,cherail,CilMD}. 
In particular, by considering the response of this BG to external nonrandom forces it was possible to establish a non-trivial fluctuation theorem and to provide explicit expressions for the effective temperatures~\cite{Lam3,Lam4}. These expressions were also recently re-examined from a different perspective in Ref.~\cite{cherail}. We also remark that this latter setting, with constant forces applied on the BG, is mathematically equivalent to the one-dimensional bead-spring model studied via Brownian-dynamics simulations in Ref.~\cite{fakhri1} and analytically in Refs.~\cite{fakhri2} and~\cite{komura}. A generalization of the BG model for a system of two coupled noisy Kuramoto oscillators, i.e., oscillators coupled by a periodic cosine potential instead  of a harmonic spring, has been discussed in Ref.~\cite{njp}.

The experiments in~Refs.~\cite{al1,al2} performed measurements of the voltages $V_1 = V_1(t)$ and  $V_2 = V_2(t)$. The spectral densities of individual realizations of $V_1$ and $V_2$, for real valued processes, are formally defined by~\cite{g1,g2} (see also~\cite{vit1,vit2}): 
\begin{equation}
\label{spectra}
\begin{split}
S_1(\omega, t) &= \frac{1}{t} \int^t_0 d\tau_1 \int^{t}_0 d\tau_2 \, \cos\left(\omega \left(\tau_1 - \tau_2\right)\right) V_1(\tau_1) V_1(\tau_2)  \,, \\
S_2(\omega, t) &= \frac{1}{t} \int^t_0 d\tau_1 \int^{t}_0 d\tau_2 \cos\left(\omega \left(\tau_1 - \tau_2\right)\right) V_2(\tau_1) V_2(\tau_2) \,,
\end{split}
\end{equation}
where the observation time $t$ is experimentally large, and mathematically intended to grow without bounds.  Note that $S_1(\omega,t)$ and $S_2(\omega, t)$ are (coupled) random functionals of a given realization of the noises $\eta_1$ and $\eta_2$, parametrized by the frequency $\omega$ (with physical dimensions $\rm rad/s$) and by the duration~$t$. 

In this work, we wish to consider two questions. First, we ask whether the statistical properties of $S_1$ and $S_2$ can discriminate between equilibrium and non-equilibrium dynamics in our system. In other words, we would like to check whether the spectral properties of the processes $V_1$ and $V_2$ for $T_1 = T_2$ differ from those for $T_1 \neq T_2$. We find that, in fact, $S_1$ and $S_2$ allow to unequivocally distinguish equilibrium from non-equilibrium. Both functionals are very simple and can be evaluated in numerical simulations or experiments, provided a good enough statistics can be gathered. Second, aiming at characterizing the statistical properties of the random functionals $S_1$ and $S_2$, we evaluate exactly their bivariate moment-generating function, defined by
\begin{equation}
\label{Phi}
\Phi(\lambda_1,\lambda_2) = \overline{\exp\left(-\lambda_1 S_1 - \lambda_2 S_2\right)} \,, 
\end{equation} 
where $\lambda_{1},\lambda_2 \geq 0$.
Given $\Phi$, we obtain the exact expression of the bivariate probability density function $P(S_1,S_2)$ by an inverse Laplace transformation. As a consequence, we are able to determine all moments and the cross-moments of $S_1$ and $S_2$ and to draw some interesting conclusions about the sample-to-sample scatter of these random functions. We emphasize that standard approaches focus exclusively on the first moments of $S_1$ and $S_2$, i.e., on the usual textbook power spectral densities. Hence, our analysis goes far beyond such a standard approach.

We obtain explicit expressions of the voltages $V_1(t)$ and~$V_2(t)$ as functions of the noises~$\eta_1(t)$ and~$\eta_2(t)$. We have indeed the following relations between the voltages $V_i(t)$ and the charges~$q_i(t)$ (see Appendix \ref{sol1}, Eqs.~\eqref{volts}):
\begin{align}
\label{voltages}
V_1 &= \frac{C+C_2}{X} q_1 - \frac{C}{X} q_2 \,, \quad
V_2 = \frac{C}{X} q_1 -  \frac{C+C_1}{X} q_2  \,. 
\end{align}
Solving Eqs.~\eqref{langevin_currents} by standard means, cf.~Appendix~\ref{sol1}, we find that for a given realization of the noises the voltages $V_1$ and  $V_2$ at time~$t$ are given by
\begin{equation}
\label{Vs1}
\begin{split}
V_1(t) &=  \frac{C+C_2}{R_1 X} \int^{t}_0 ds \, Q(a_1, t - s) \eta_1(s)   \\ &-\frac{C}{R_2 X}   \int^{t}_0 ds \, Q(v, t - s) \eta_2(s)\,, \\
V_2(t) &= \frac{C}{R_1 X} \int^{t}_0 ds \, Q(v, t - s) \eta_1(s)  \\&- \frac{C + C_1}{R_2 X}   \int^{\tau}_0 ds \, Q(a_2, t - s) \eta_2(s)\,.
\end{split}
\end{equation}
Here the kernel $Q$ is given by
\begin{align}
\label{Q}
Q(a,s) = \exp(- v s) \left[\cosh\left(b s\right) - \frac{a}{b} \sinh\left(b s\right)\right] \,,
\end{align}
while the parameters $v$, $b$, $a_1$ and $a_2$ obey
\begin{equation}
\label{param}
\begin{split}
v & = \frac{R_1 (C + C_1) + R_2 (C + C_2)}{2 R_1 R_2 X} \,,  \\
b &= \sqrt{v^2 - \frac{1}{R_1 R_2 X}} \,, \\
a_1 &= \frac{R_2 \left(C+C_2\right)^2 - R_1 \left(X - C^2\right)}{2 \left(C+C_2\right) R_1 R_2 X} \,, \\
a_2 &= \frac{R_1 (C + C_1)^2 - R_2  \left(X - C^2\right)}{2 \left(C+C_1\right) R_1 R_2 X} \,.
\end{split}
\end{equation}
In the following we take advantage of these explicit expressions to answer the two questions raised above.

\section{In or out of equilibrium? Mean and correlation of single-trajectory spectral densities}
\label{main}

In this section, we consider the first moments of $S_{i}(\omega,t)$, $i=1,2$, in the limit $t\to\infty$, i.e., the standard power spectral densities~$\mu_i(\omega)$ of the measured voltages $V_1$ and $V_2$. We define therefore, for $i=1,2$,
\begin{align}
\label{nu}
S_i(\omega)&=\lim_{t\to\infty}S_i(\omega,t);\\
\label{nu2}
\mu_i(\omega)&=\overline{S_i(\omega)}.
\end{align}
We show below that the knowledge of the exact expressions of these moments is important for two reasons: on the one hand, the higher-order moments of the $S_i(\omega)$ can all be expressed in terms of the~$\mu_i(\omega)$; and on the other hand, the single-trajectory spectral densities take the form $S_i(\omega) = r_i \, \mu_i(\omega)$, $i=1,2$, in the limit of an infinitely long observation time, where the $r_i$s are random, exponentially distributed dimensionless amplitudes. Hence, the entire $\omega$-dependence, as well as the dependence on the temperatures and the material parameters is fully encoded in the first moments. We show that it is possible to introduce an observable related to the first moments that helps in discriminating equilibrium from non-equilibrium.
We also discuss the Pearson coefficient of the spectral densities~$S_1(\omega)$ and~$S_2(\omega)$ and show that it also allows to distinguish equilibrium from non-equilibrium.

\subsection{Explicit expressions of the first moments of the $S_i(\omega)$}
We derive in Appendix~\ref{B} the following explicit expressions of the first moments of $S_1(\omega,t)$ and $S_2(\omega, t)$ in the limit $t \to \infty$, (see Eq. \ref{nu2}): 
\begin{equation}
\label{eq:mean_spectra}
\begin{split}
\mu_1(\omega) & = \frac{4 k_\mathrm{B}}{X^2 \left(b^4 + 2 b^2 (\omega^2 - v^2) + (\omega^2 + v^2)^2\right)}  \\ 
&\qquad \times \Bigg(\Bigg[ \frac{(C+C_2)^2 \left( \omega^2 + (v-a_1)^2 \right)}{R_1} + \frac{C^2 \omega^2}{R_2}\Bigg] T_1  
\\
&\qquad + \frac{C^2 \omega^2}{R_2} (T_2 - T_1)\Bigg) \,, \\
\mu_2(\omega)  &= \frac{4 k_\mathrm{B}}{X^2 \left(b^4 + 2 b^2 (\omega^2 - v^2) + (\omega^2 + v^2)^2\right)}  \\ 
&\qquad \times \Bigg( \Bigg[\frac{(C+C_1)^2 \left(\omega^2 + (v-a_2)^2\right)}{R_2} +  \frac{C^2 \omega^2}{R_1}\Bigg] T_2  \\
&+ \frac{C^2 \omega^2}{R_1} (T_1-T_2)\Bigg) \, .
\end{split}
\end{equation}
Equations~\eqref{eq:mean_spectra} have been previously obtained in Refs.~\cite{al1,al2}, by applying the fluctuation-dissipation theorem directly to the circuit defined by Eqs. \eqref{langevin_currents}. For finite values of~$t$ there are corrections of order~$1/t$. Therefore, numerical or experimental tests require going to sufficiently long times. However, the experimentally relevant values of the system parameters imply very small characteristic relaxation times, and the time-dependent corrections, which we evaluate explicitly in Appendix~\ref{C}, become negligibly small already at times of the order of seconds. Hence, the asymptotic forms in Eqs. \eqref{eq:mean_spectra} can be readily accessed experimentally.

\begin{figure}[htbp]
	\begin{center}
		\includegraphics[width=0.47\textwidth]{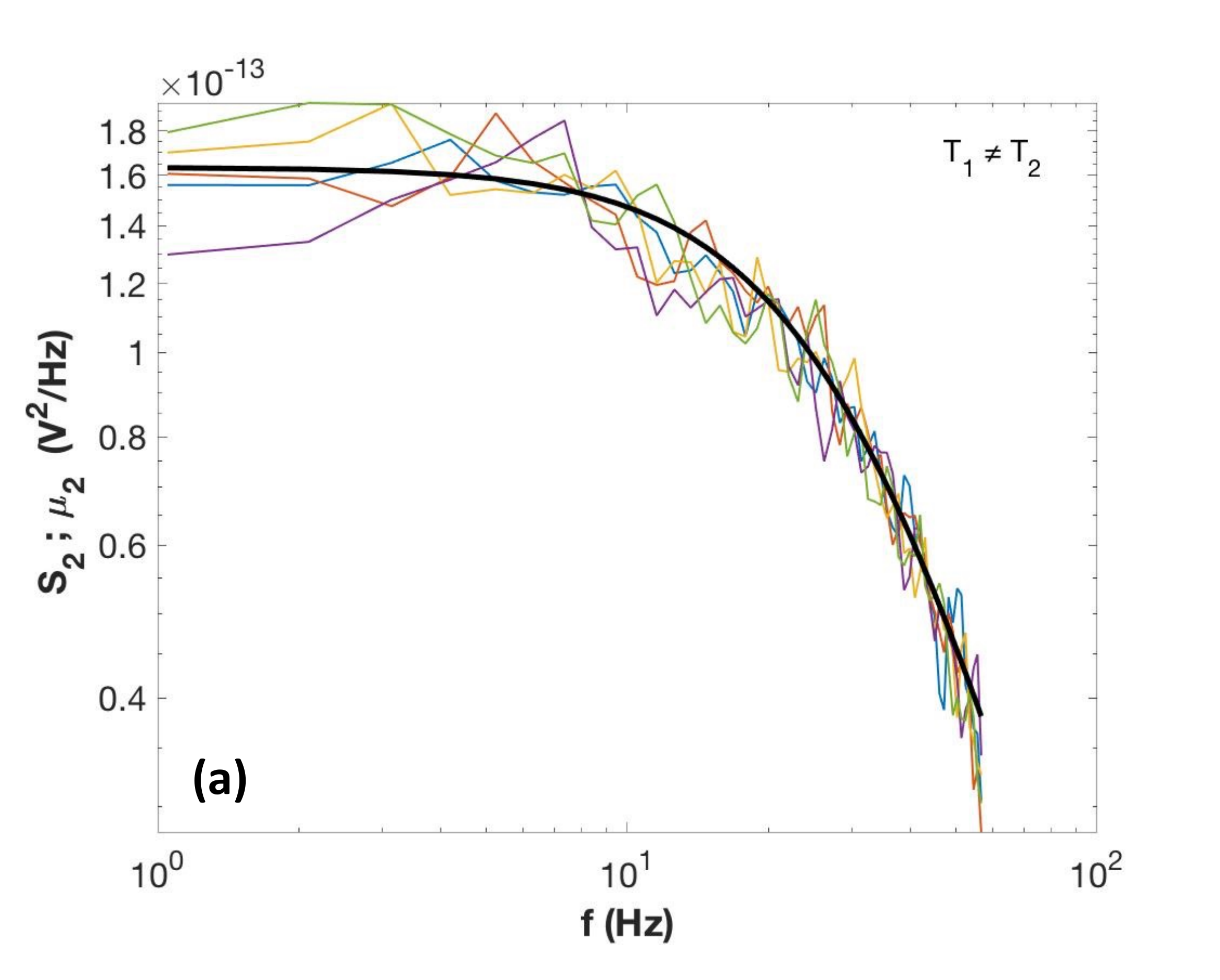}\hspace{0.05\textwidth}
		\includegraphics[width=0.47\textwidth]{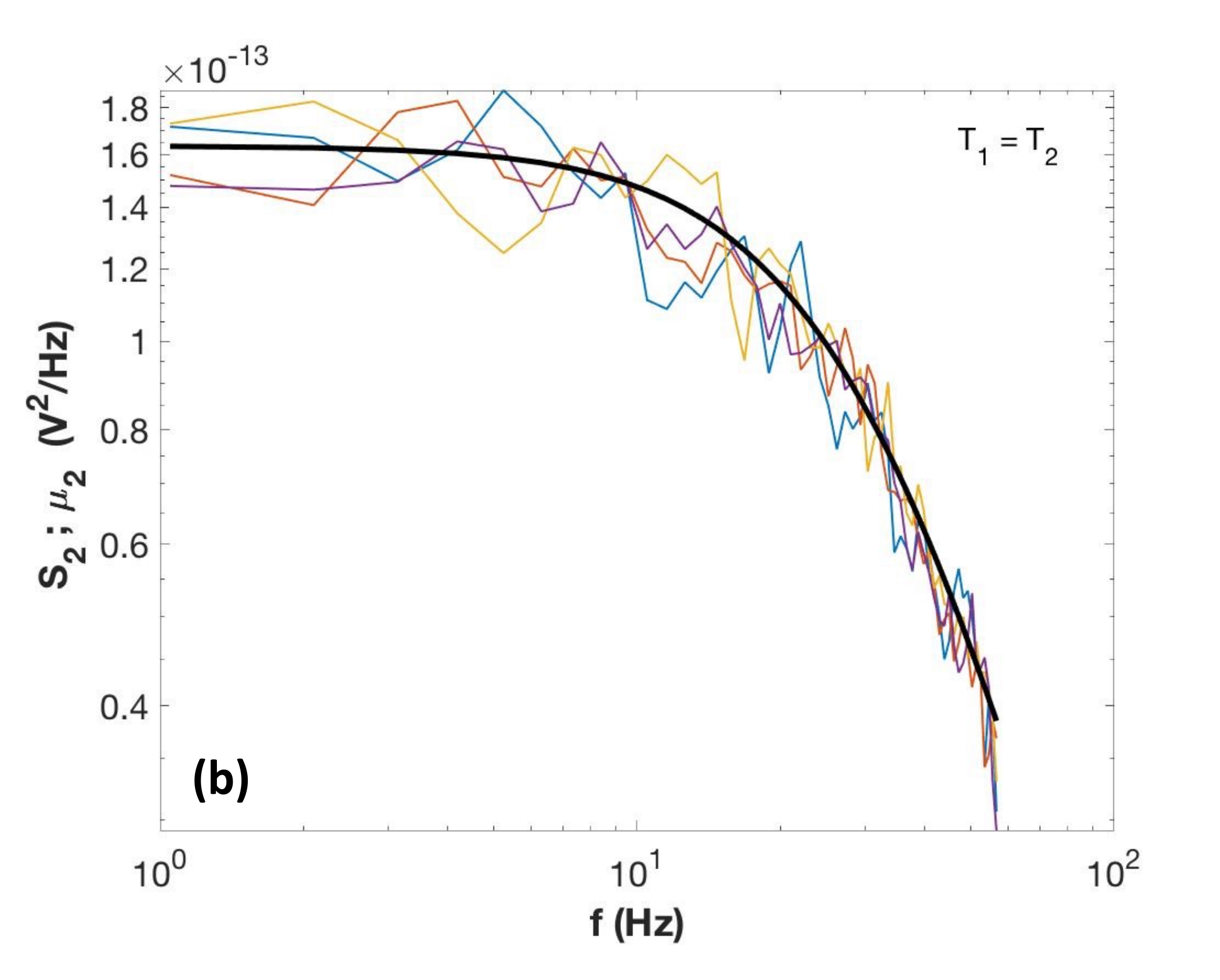}
	\end{center}
	\caption{(Color online) Ensemble-averaged power-spectral density $\mu_2(\omega)$ (thick solid curve), Eqs. \eqref{eq:mean_spectra}, as a function of the frequency $f=\omega/2 \pi$, together with realization-dependent random function $S_2(\omega)$ (thin noisy curves) for five randomly chosen realizations of voltages recorded in experiments in Refs.~\cite{al1,al2}. Panel (a): out-of-equilibrium dynamics with $T_1 = 88 \, \unit{K}$ and  $T_2 = 296 \, \unit{K}$. Panel (b): equilibrium dynamics with $T_1 = T_2 = 296 \, {\rm K}$. Material parameters are as defined in the caption to Fig. \ref{fig3}.}
	\label{fig2}
\end{figure}
 
In Fig.~\ref{fig2} we plot the ensemble-averaged spectral density $\mu_2(\omega)$ as a function of $\omega$, together with five realizations of the random function $S_2(\omega)$ evaluated  for five individual realizations of the experiment~~\cite{al1,al2}. (The behavior of~$\mu_1(\omega)$ is similar.) We observe that the random functions $S_2$ follow rather closely the average curve $\mu_2(\omega)$ for $\omega>0$. In the vicinity of $\omega =0$ the scatter is more significant.  We discuss in Sec.~\ref{dist} the reasons of this behavior. Importantly, since the random functions $S_2(\omega)$ lie close to their ensemble-averaged value $\mu_2(\omega)$, it appears that a modest statistics already allows to evaluate~$\mu_2(\omega)$ reliably  well from experimental data.

\subsection{Discriminating equilibrium from non-equilibrium}
In their leading order in the large $\omega$ regime $\mu_1(\omega)$ and $\mu_2(\omega)$ decrease like~$\omega^{-2}$:
\begin{equation}
\label{z}
\begin{split}
&\mu_1(\omega) = \frac{4 k_\mathrm{B}}{X^2 \, \omega^2} \Bigg(\frac{(C+C_2)^2}{R_1} T_1 + \frac{C^2}{R_2} T_2\Bigg) + O\left(\frac{1}{\omega^4}\right)\,,\\
&\mu_2(\omega) = \frac{4 k_\mathrm{B}}{X^2 \, \omega^2} \Bigg(\frac{C^2}{R_1} T_1 + \frac{(C+C_1)^2}{R_2} T_2\Bigg) + O\left(\frac{1}{\omega^4}\right) \,.
\end{split}
\end{equation}
This means that the short-time dynamics, corresponding to large $\omega$, is Brownian. The noise amplitudes of this dynamics depend on all material parameters and on both temperatures, but in cases with $T_1 \neq T_2$ they 
do not sensibly differ from cases with $T_1 = T_2$. 
On the other hand, their maximum values, corresponding to the zero frequency 
limit (hence, to the long time behavior), take the very simple form:
\begin{equation}
\label{zz}
\mu_i(0) = \lim_{t \to \infty} \frac{1}{t} \overline{\left(\int^t_0 d\tau V_i(\tau)\right)^2} = 4 k_\mathrm{B} R_i T_i  \,,\qquad i=1,2\,. \\
\end{equation}

From a mathematical point of view, $\mu_1(0)$ and $\mu_2(0)$ are the $t\to\infty$ limit of the averaged squared areas under the random curves $V_1$ and $V_2$, respectively, divided by the observation time $t$. One also observes that $x_1 = \int^t_0 d\tau V_1(\tau)$ and $x_2 = \int^t_0 d\tau V_2(\tau)$ can be thought 
of as instantaneous positions of two particles which perform diffusive motion with the "diffusion coefficients" $2 k_\mathrm{B} R_1 T_1$ and $2 k_\mathrm{B} R_2 T_2$, respectively.
These "diffusion coefficients" depend only on the temperature and on the resistance of the component under consideration. In particular, they neither depend on the temperature nor on the resistance of the second component, nor on other material parameters.  
\textit{A priori}, this compensation of dependencies in the asymptotic large-$t$ limit is rather unexpected. However, we show in Appendix~\ref{C} that the covariance function of $S_1$ and $S_2$ vanishes at $\omega = 0$, so that the evolution of the area under one of the components indeed decouples at long times
from the evolution of the second component.

Therefore, if the precise values of the parameters of the circuit are known, we can estimate the temperatures~$T_1$ and~$T_2$ from the experimental data by means of~Eqs.~\eqref{eq:mean_spectra} and~\eqref{zz}, if a sufficiently large statistics is available. However, a better candidate for such an analysis is provided by the spectral density evaluated at $\omega=0$, since it only requires knowledge of the resistances $R_1$ and $R_2$,  and does not necessitate the knowledge of the values of the capacitances. We define therefore the random variable~$\chi_t$ by
\begin{equation}\label{chi}
\chi_t=\frac{1}{k_\mathrm{B}} \left(\frac{S_1(\omega=0,t)}{R_1}-\frac{S_2(\omega=0,t)}{R_2}\right),
\end{equation}
whose average for large values of~$t$ is given by
\begin{align}
\label{Rparameter}
\overline{\chi} =\frac{1}{k_\mathrm{B}} \left(\frac{\mu_1(0)}{R_1} - \frac{\mu_2(0)}{R_2}\right) = 4 \left( T_1 - T_2 \right) \,.
\end{align}
This average, which has the dimensions of temperature and does not depend on the material parameters $R_1$ and $R_2$,
allows to discriminate equilibrium from non-equilibrium, since it
vanishes if and only if $T_1 = T_2$.  The finite time expression \eqref{chi}, depends on $R_1$ and $R_2$, but for
$R_1=R_2$ the exact value of the resistance becomes equally irrelevant to discriminate between equilibrium and non-equilibrium
situations. 
This is the case, for instance, in the experimental realization of a Brownian gyrator model made by a single colloidal particle subject to an elliptical confining potential~\cite{11}. In this case $R_1 =R_2 = R$, where $R$ is the 
isotropic Stokes coefficient, which is the same for both components. In the completely symmetric case, where not only $R_1 = R_2$ but also $C_1 = C_2 = C$, one can explicitly evaluate the full dependence of~$\overline{\chi}$ on the observation time~$t$ (see Eq. \eqref{arbchi}). In other kinds of experiments, and in particular in those of Refs.~\cite{al1,al2}, this is not the case.

\subsection{Pearson coefficient of the spectral densities}
We now turn our attention to the Pearson correlation coefficient of the single-trajectory spectral densities $S_1$ and $S_2$,  defined by
\begin{equation}
\label{pearson}
\rho_{\omega} = \lim_{t \to \infty} 
\frac{\overline{
\left(S_1(\omega) - \mu_1(\omega)\right) \left(S_2(\omega) - \mu_2(\omega)\right)}}{\sqrt{\left(\overline{S_1(\omega)^2} - \mu^2_1(\omega)\right) \left(\overline{S_2(\omega)^2} - \mu^2_2(\omega)\right)}} \,,
\end{equation} 
where the numerator is the covariance of the random functionals $S_1(\omega)$ and $S_2(\omega)$, while the denominator is the product of their standard deviations. To our knowledge, this quantity has not been previously considered in the present context. As shown in Appendix~\ref{C}, $\rho_{\omega}$ is given by:
\begin{widetext}
\begin{align}
\label{pearson2}
\rho_{\omega} = \frac{R_1 R_2 C^2 \omega^2  \left(\left(T_1 - T_2\right)^2 + \omega^2 \left[(C+C_2) R_2 T_1 + (C+C_1) R_1 T_2\right]^2\right)}{\left(T_1 + R_2 \omega^2 \left[(C+C_2)^2 R_2 T_1 + C^2 R_1 T_2\right]\right) \left(T_2 + R_1 \omega^2 \left[(C+C_1)^2 R_1 T_2 + C^2 R_2 T_1\right]\right)} \,.
\end{align}
\end{widetext}
Clearly, $\rho_{\omega}$ is positive and bounded from above by unity, and vanishes only for $\omega =0$, further confirming that the squared areas under the random curves $V_1$ and $V_2$ become statistically independent in the $t \to \infty$ limit.  The dependence of $\rho_{\omega=0}$ on the observation time~$t$ is analyzed in Appendix~\ref{C}, where it is shown that at large times $\rho_{\omega=0} = O(1/t^2)$ (see Eq. \eqref{rho0as}). For fixed $\omega >0$ and $t \to \infty$, the Pearson coefficient defined by~Eq. \eqref{pearson2} attains its maximal value $\rho_{\omega}= 1$ when either 1) one temperature is finite and the other is infinitely large, or 2) one temperature is 
finite and the other vanishes. The second is a very peculiar case, in which the dynamics of the ``passive'' zero temperature component is essentially enslaved
to the non-zero temperature ``active'' component (see \cite{Lam3} for more details).   A direct consequence of points 1) and 2) is that $\rho_{\omega} $ is a \textit{non-monotonic} function of the temperatures. Indeed, suppose that one keeps~$T_2$ fixed and varies $T_1$, as in the experiments in \cite{al1,al2}: one then has $\rho_{\omega} = 1$ both for~$T_1 = 0$ and~$\rho_{\omega} = 1$ for $T_1 \to \infty$, which implies that there exists a temperature $T_1^{*}$ at which the correlations between the spectral densities of both components are minimal. This value of the temperature can be readily found from Eq. \eqref{pearson2}:
\begin{equation}
T_1^{*} = \left(\frac{1 + \left(C+C_1\right)^2 R_1^2 \omega^2}{1 + \left(C+C_2\right)^2 R_2^2 \omega^2}\right)^{1/2} T_2 \,.
\end{equation}
Somewhat counter-intuitively, $T_1^{*} \neq T_2$, i.e., the smallest correlation appears in the non-equilibrium case.

Consider next the limiting behavior of $\rho_{\omega}$. In the large-$\omega$ limit, one has
\begin{equation}
\begin{split}
\rho_\infty&= \lim_{\omega \to \infty} \rho_{\omega} = C^2 \left[(C+C_2) R_2 T_1 + (C+C_1) R_1 T_2 \right]^2 \\
& \qquad \times\left[(C+C_1)^2 R_1 T_2 + C^2 R_2 T_1\right]^{-1} \\& \qquad \times \left[(C+C_2)^2 R_2 T_1 + C^2 R_1 T_2\right]^{-1}  \,,
\end{split}
\end{equation}
which becomes temperature independent if $T_1 = T_2$. However, it is rather difficult to access the large-$\omega$ regime in practice, because dealing with finite data sets, $S_1$ and $S_2$ appear as periodic functions of $\omega$, and some care has to be taken in fitting the data  \cite{g1,g2}. In contrast, the small-$\omega$ regime is relatively easier to access. In this limit, and for $T_1$ and  $T_2$ both bounded away from zero, we have
 \begin{equation}
 \label{asymp}
 \rho_{\omega}  = C^2 R_1 R_2 \frac{\left(T_1 - T_2\right)^2}{T_1 T_2} \omega^2 + A(T_1,T_2) \omega^4 + O\left(\omega^6\right) \,,
 \end{equation}
where $A(T_1,T_2)$ depends on the temperatures and on the material parameters. Then, in the small $\omega$ regime, $\rho_{\omega} \sim \omega^4$ for $T_1 = T_2$, while $\rho_{\omega} \sim \omega^2$ for $T_1 \ne T_2$.  

This allows us, in fact, to distinguish equilibrium from non-equilibrium dynamics by concentrating on the small-$\omega$ asymptotic behavior of the Pearson correlation coefficient of the spectral densities of two components, \textit{without any knowledge of material parameters} --- the dependence on the frequency is strikingly different in equilibrium and in non-equilibrium. In Fig.~\ref{fig3} we present a comparison of our theoretical prediction in Eq. \eqref{pearson2} with experimental data and  numerical simulations, which shows that such an analysis is indeed possible. We remark, however, that evaluating~$\rho_{\omega}$ requires a much larger statistical sample than the analysis of the ensemble-averaged power spectral densities, since it involves the fluctuations around the mean.

\section{Moment-generating and probability density functions of single-trajectory spectral densities}
\label{dist}
We now focus on the statistical properties of spectral densities of individual realizations of $V_1$ and $V_2$. To this end, we evaluate exactly the bivariate moment-generating function $\Phi(\lambda_1,\lambda_2)$, defined in~\eqref{Phi}, and we then evaluate the joint bivariate probability density function of the spectral densities of both components by inverting the Laplace transform.  Since the limits $t \to \infty$ and $\omega \to 0$ do not commute, in accord with previous findings made for unbounded Gaussian processes \cite{g1,g2,vit1,vit2}, we consider separately the behavior at $t = \infty$ for arbitrary $\omega > 0$, and the one at~$\omega = 0$ for arbitrary $t$.  The results obtained for the latter case allow us to evaluate the full probability density function of the observable~$\chi$ defined in~Eq. \eqref{chi}. Below we merely list our main results. More discussions and the (rather lengthy) calculations are presented in Appendix~\ref{C}.

\subsection{The limit $t \to \infty$ for $\omega > 0$}
In the $t \to \infty$ limit, the moment-generating function~$\Phi$ has the expression
\begin{align}
\label{mg}
\Phi(\lambda_1,\lambda_2) &= \Big[\left(1 +  \mu_1(\omega) \lambda_1\right)\left(1+ \mu_2(\omega) \lambda_2\right) \nonumber\\ \qquad &-
 \rho_{\omega} \mu_1(\omega) \mu_2(\omega) \lambda_1 \lambda_2\Big]^{-1} \,,
\end{align}
where $\mu_1(\omega)$ and $\mu_2(\omega)$ are defined in Eq. \eqref{eq:mean_spectra}, while the Pearson coefficient $\rho_{\omega}$ is defined by Eq.~\eqref{pearson2}. Differentiating $\Phi$ with respect to to $\lambda_1$ and $\lambda_2$ yields the moments and cross-moments of the spectral densities, while their probability density function is obtained from the inverse Laplace transform of $\Phi$.

\subsubsection{Bivariate probability density function}
Performing the inverse Laplace transform of the moment-generating function $\Phi$, we obtain the bivariate probability density function $P(S_1=s_1,S_2=s_2)=P(s_1,s_2)$:
\begin{equation}
\label{joint2}
\begin{split}
P(s_1,s_2)
&=\dfrac{1}{\mathcal{N}} \exp\left(-\frac{s_1}{\left(1 - \rho_{\omega}\right) \mu_1(\omega)} - \frac{s_2}{\left(1 - \rho_{\omega}\right) \mu_2(\omega)}\right) \\
&\qquad {}\times I_0\left(\frac{2}{1 - \rho_{\omega}} \sqrt{\frac{\rho_{\omega} s_1 s_2}{\mu_1(\omega) \mu_2(\omega)}}\right) \,,
\end{split}
\end{equation}
where $\mathcal{N}=\mu_1(\omega) \mu_2(\omega) \left(1 - \rho_{\omega}\right)$ is a normalizing factor and~$I_0$ is the modified Bessel function of the first kind. When one of the temperatures vanishes, say $T_1 \to 0$ while $T_2 > 0$, $\rho_{\omega}$ tends to $1$, and the Bessel function in~Eq. \eqref{joint2} approaches the Dirac delta function $\delta\left(s_1 - s_2\right)$. 

If we fix~$s_2$ and vary~$s_1$, we realize that there are two different regimes for $P(s_1,s_2)$, depending on the value of~$s_2$. For $s_2 < s_2^\mathrm{th}(\omega)$, where the threshold $s_2^\mathrm{th}(\omega)$ is given by
\begin{equation}
\label{S2threshold}
s_2^\mathrm{th}(\omega) = \frac{\mu_1(\omega) \left(1- \rho_{\omega}\right) }{\rho_{\omega}}  \,,
\end{equation}
$P(s_1,s_2)$ is a monotonically decreasing function of $s_1$, with a maximum at $s_1 = 0$. Otherwise, for $s_2 > s_2^\mathrm{th}(\omega)$, $P(s_1,s_2)$ has a maximum at a value $s_1^* > 0$, meaning that there exists a most probable value of the spectral density of the first component. Note that the threshold $s_2^\mathrm{th}(\omega)$ depends on the frequency, and therefore that for sufficiently low frequencies the probability density function can be non-monotonic as a function of $s_1$, while becoming monotonic for higher frequencies.

The moments and the cross-moments 
of any, not necessarily integer, orders $n$ and $m$ can be explicitly evaluated from~Eq.~\eqref{joint2}.
Multiplying Eq. \eqref{joint2} by $s_1^n s_2^m$ and integrating, we obtain
\begin{equation}
\begin{split}
\overline{S_1^n S_2^m} &= \Gamma(n+1) \Gamma(m+1) \mu_1^n(\omega) \mu_2^m(\omega) \left(1 - \rho_{\omega}\right)^{n+m+1}\\
&\qquad \times _2F_1\left(n+1,m+1,1;\rho_{\omega}\right) \,,
\end{split}
\end{equation}
where $\,_2F_1$ is the Gauss hypergeometric function.

\subsubsection{Univariate probability density function}
The univariate moment-generating function of, say, $S_1$ 
is obtained from Eq.~\eqref{mg} by simply setting $\lambda_2$ to~0, which yields
\begin{equation}
\label{phix3}
\Phi(\lambda_1) = \frac{1}{1 +  \mu_1(\omega)  \lambda_1 } \,.
\end{equation}
This expression corresponds to a simple exponential distribution for $S_1$:
\begin{equation}
\label{distS}
P(s_1) = \dfrac{1}{\mu_1(\omega)}\exp\left(- \dfrac{s_1}{\mu_1(\omega)}\right) \,,
\end{equation}
with a maximum at $s_1 = 0$. 
 
\begin{figure}[htp]
	\begin{center}
		\includegraphics[width=0.48\textwidth]{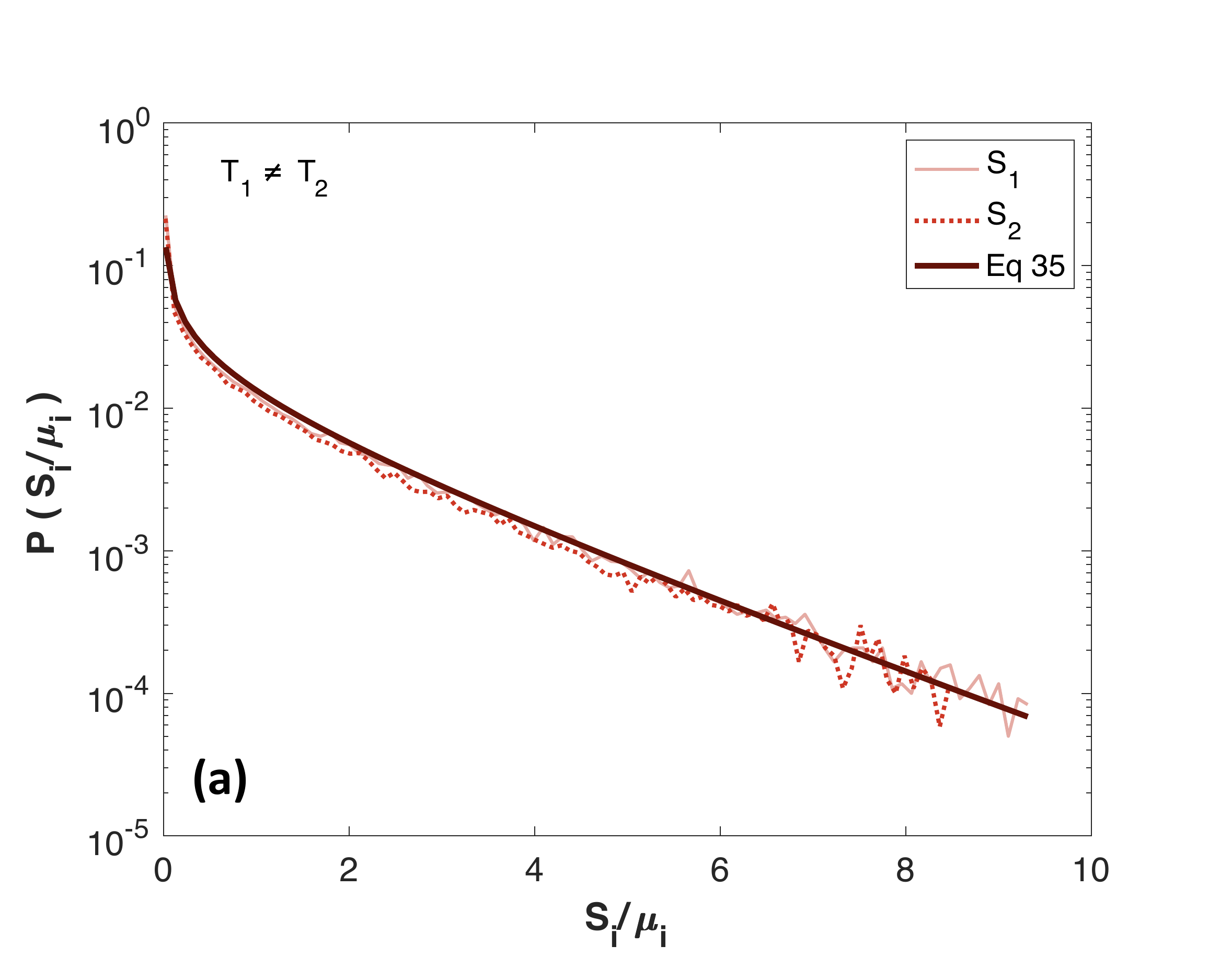}
		\includegraphics[width=0.48\textwidth]{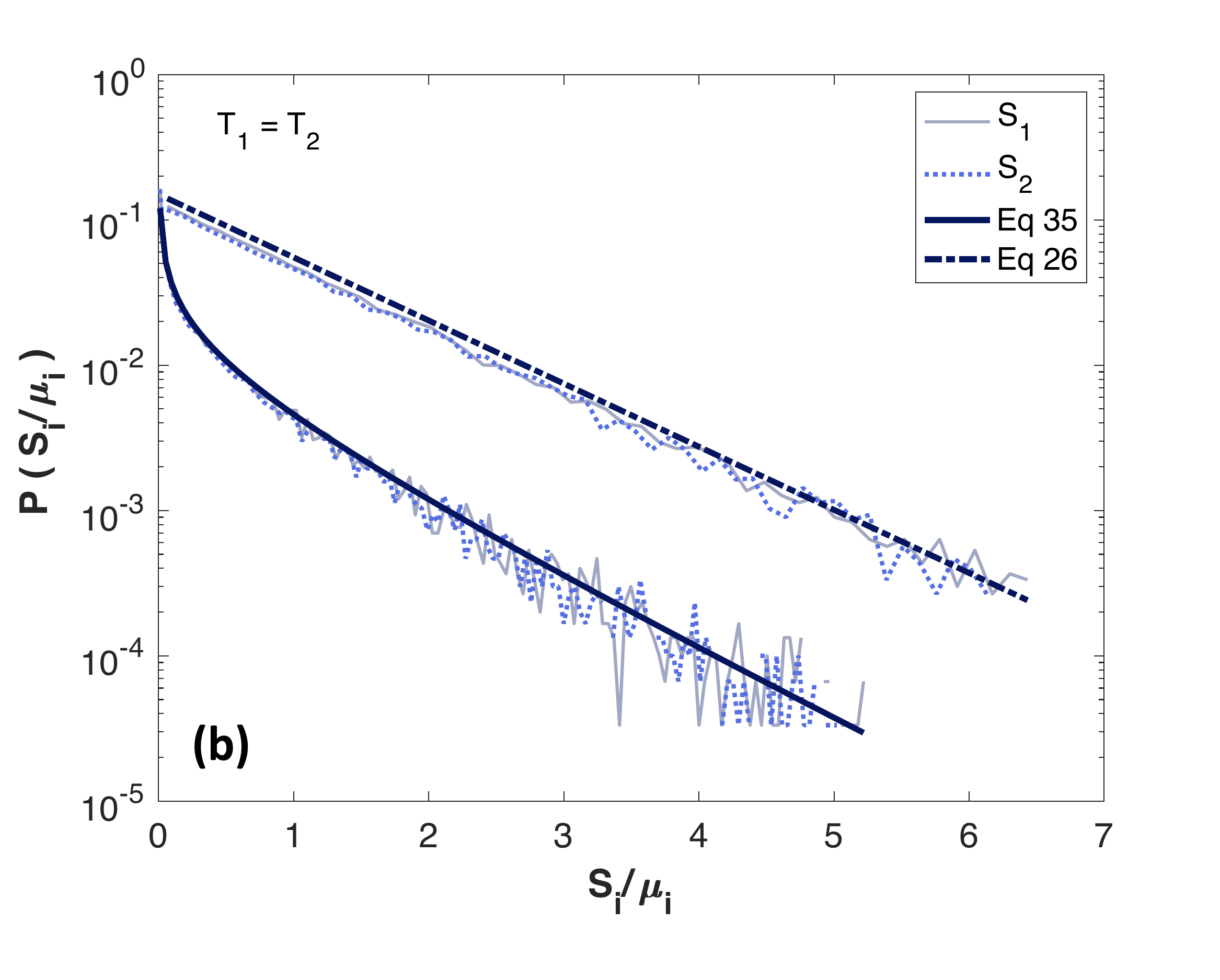}
	\end{center}
	\caption{(Color online) The univariate probability density functions in Eqs. \eqref{distS} and \eqref{P3}. Panel (a): Distribution of single trajectory spectra out of equilibrium for $\omega = 0$ ($f=0$). Theoretical prediction in Eq. \eqref{P3} (thick solid curve) vs. experiment (same parameters as~in Ref.~\cite{al2} and in the caption to Fig. \ref{fig3}). $S_1$ (light red curve): component~$x$, $S_2$ (dark red curve).  
	Panel (b): Distributions of single trajectory spectra at equilibrium. Theoretical predictions (dash-dotted curve)
	in Eq. \eqref{distS} (upper curves, $f = 1 \, \unit{Hz}$) and in Eq. \eqref{P3} (lower solid curves, $f=0$) vs. experiment in Ref.~\cite{al2} (same parameters). $S_1$ (gray curves): component~$x$, $S_2$ (light blue curves): component~$y$. Note that a slight change of the value of the frequency ($f= 1 \, \unit{Hz}$ vs $f =0$) results in a drastic change of the shape of the probability density function, in accord with our theoretical predictions.} 
	\label{fig4}
\end{figure}

In Fig.~\ref{fig4}  we present the comparison of our predictions in Eqs. \eqref{distS} and \eqref{P3} against the experimental data, which shows that our theoretical results are fully corroborated by the experimental data.
Further, 
the moments of $S_1$ of any real order $k > 0$ can be obtained from Eq.~\eqref{distS} and have a very simple form:
\begin{equation}
\overline{S_1^k} = \Gamma(k+1) \mu_1^k(\omega) \,.
\end{equation}
In turn, the typical value $S_1^\mathrm{typ}$, defined by~\cite{typ}
\begin{equation}
\label{typ}
S_1^\mathrm{typ} =  \exp\left(\overline{\ln S_1}\right)\, ,
\end{equation} 
is given by
\begin{equation}
S_1^\mathrm{typ} = \mu_1(\omega) \, e^{-\gamma_{\mathrm{E}}} \,,
\end{equation}
where $\gamma_{\mathrm{E}} \approx 0.577$ is the Euler gamma constant.  

Let us make some further comments on Eqs.~\eqref{distS} to~\eqref{typ}:
\begin{itemize}
\item Eq.~\eqref{distS} to Eq.~\eqref{typ} show that the entire dependence of the power spectral density $S_1$ on $\omega$, (and similarly, of $S_2$ on $\omega$), for a given realization of noises, 
is encoded in the ensemble-averaged value $\mu_1(\omega)$ (or $\mu_2(\omega)$). This implies that Eq.\eqref{distS} can be rewritten as an equality in distribution:
\begin{equation}
\label{r}
\frac{S_1}{\mu_1(\omega)} \overset{\mathrm{d}}{=} r \,,
\end{equation}
where $r$, for each frequency $\omega$, is a random dimensionless variable with a universal distribution $P(r) = \exp(- r)$.  As shown in \cite{newest}, variables corresponding to different frequencies are independent. Therefore, the form of the ensemble-averaged power spectrum density can
in principle be inferred from a single, albeit long, realization of $V_1$, without any reference to ensemble averaging. See~Refs.~\cite{g1,g2,vit1,vit2} for a more ample discussion of this point in case of unbounded Gaussian processes. \item The probability density function in Eq.~\eqref{distS} is \textit{effectively} broad (see Appendix~\ref{D} for more details). Indeed, the standard deviation $\sigma_1(\omega)$ of the single-trajectory power spectral density $S_1$ is equal to its mean $\mu_1(\omega)$, 
\begin{equation}
\label{moments11}
\sigma_1(\omega) = \mu_1(\omega) \,, \quad \omega > 0 \,.
\end{equation}
In other words, the coefficient of variation $\mathsf{cv}(\omega) = \sigma_1(\omega)/\mu_1(\omega)$ of the distribution in Eq.~\eqref{distS}  is exactly equal to unity. 
\item  Interestingly enough,  $S_1^\mathrm{typ}$ in Eq.~\eqref{typ}  depends on~$\omega$ exactly
as its ensemble-averaged counterpart does, but has a smaller amplitude. This suggests that the ensemble-averaged property~$\mu_1(\omega)$ is supported by some \textit{atypical} realizations of $V_1$.  
\end{itemize}

\subsection{The limit $\omega = 0$ for arbitrary $t$}
As shown in Appendix~\ref{C}, the joint moment-generating function $\Phi(\lambda_1,\lambda_2)$ for $\omega =0$ and arbitrary $t$ is given by (see Appendix \ref{C} for the details of the derivation)
\begin{align}
\label{Phif0}
\Phi(\lambda_1,\lambda_2) &= \Bigg[\left(1 + 2 \, \Sigma_1(t) \, \lambda_1\right) \left(1 + 2 \,\Sigma_2(t) \, \lambda_2\right) \nonumber\\ \qquad &- 4 \rho_0 \,  \lambda_1 \,\lambda_2\, \Sigma_1(t) \, \Sigma_2(t)\Bigg]^{-1/2} \,,
\end{align}
where $\Sigma_1(t)=\overline{S_1(\omega=0,t)}$ and $\Sigma_2(t)=\overline{S_2(\omega=0,t)}$ are the ensemble-averaged spectral densities of $V_1$ and $V_2$ calculated at zero frequency, 
i.e., the averaged squared areas below $V_1$ and $V_2$ divided by $t$, while $\rho_0$ is the $t$-dependent, $\omega=0$ Pearson coefficient (see Eq.~\eqref{pearson4}). The functions 
$\Sigma_1(t)$ and $\Sigma_2(t)$ are monotonically increasing functions of $t$, which approach for $t \to \infty$ their asymptotic values $4 k_\mathrm{B} R_1 T_1$ and $4 k_\mathrm{B} R_2 T_2$, respectively, while $\rho_0 \to 0$ in this limit.
For arbitrary $t$, the exact expressions for 
$\Sigma_1(t)$, $\Sigma_1(t)$ and $\rho_0$  are rather lengthy and we do not present their explicit forms. In Appendix~\ref{C} we show that they approach their limiting values as $\Sigma_i(t) = 4 k_\mathrm{B} R_i T_i + O(1/t)$, $i=1,2$, while $\rho_0 = O(1/t^2)$. 

\subsubsection{Bivariate probability density function}
The expression in Eq.~\eqref{Phif0}  is functionally different from the result in Eq.~\eqref{mg}, corresponding to the limit $t \to \infty$. We can therefore expect that the bivariate and univariate probability density functions will be different from the expressions appearing in Eqs.~\eqref{joint2} and \eqref{distS}.  Expanding the expression \eqref{Phif0} in powers of the Pearson coefficient $\rho_0$, inverting the Laplace transform and summing the resulting series, we obtain
\begin{equation}
\label{dist_voltages}
\begin{split}
&P(s_1, s_2) = \frac{1}{2 \pi\sqrt{(1-\rho_0)\Sigma_1(t)\Sigma_2(t)s_1s_2}} \\ &\qquad \times \cosh\left(\frac{\sqrt{\rho_0}}{1-\rho_0} \sqrt{\dfrac{s_1}{\Sigma_1(t)} \, \dfrac{s_2}{\Sigma_2(t)}} \right) \\
& \qquad \times \exp\left(- \dfrac{s_1}{2 \left(1 - \rho_0\right) \Sigma_1(t)} - \dfrac{s_2}{2 \left(1 - \rho_0\right)\Sigma_2(t)}\right) \,.
\end{split}
\end{equation}
This is the desired bivariate distribution of the squared areas under the random curves $V_1$ and $V_2$, divided by the arbitrary observation time $t$, which indeed has a different form from that given in~Eq. \eqref{joint2}. In the limit $t \to \infty$ the Pearson coefficient $\rho_0$ vanishes, so that the cosh factor in~Eq. \eqref{dist_voltages} becomes equal to $1$. As a consequence, the probability density function factorizes into the product of two exponentials. Moreover,
the bivariate distribution of the integrated voltages themselves can be derived from Eq.~\eqref{dist_voltages} by a simple change of variables. In the limit $t \to \infty$ it becomes a product of two statistically independent Gaussian functions. Last, we remark that some qualitative features of the distribution in~Eq. \eqref{dist_voltages} resemble those of the one in~Eq. \eqref{joint2}. Indeed, it is a non-monotonic function of $s_1$ when $s_2$ exceeds some time-dependent threshold value, and it is a monotonically decreasing function of $s_1$, otherwise.

\subsubsection{Univariate probability density function}
The univariate distribution of a single component, e.g., $V_1$, is obtained by setting $\lambda_2 = 0$ in Eq.~\eqref{Phif0}. Thus the moment generating function of $S_1(\omega=0,t)$ is given by
\begin{equation}
\label{Phif01}
\Phi(\lambda_1) = \frac{1}{\sqrt{1 + 2 \, \Sigma_1(t) \, \lambda_1}} \,,
\end{equation}
and therefore the univariate probability density function is given by
\begin{equation}
\label{P3}
P(S_1(\omega=0,t) = s_1) = \frac{1}{\sqrt{2 \pi  \Sigma_1(t) s_1}} \exp\left(- \dfrac{s_1}{2 \Sigma_1(t)}\right).
\end{equation}
The probability density function in~Eq. \eqref{Phif0} is shifted to the left, favoring very small values of $s_1$, as compared to the one in~Eq. \eqref{distS}. Indeed,  $P(S_1(\omega=0,t) = s_1)$ diverges when $s_1 \to 0$, (while the probability density function in Eq.~\eqref{distS} remains finite in this limit), and its
right tail decreases somewhat faster due to an additional factor $1/\sqrt{s_1}$. 

From Eq. \eqref{P3} we obtain the moments of order~$k$ (not necessarily integer), for $k > - 1/2$:
\begin{equation}
\label{nn1}
 \overline{S^k_1(\omega=0,t)} = \frac{2^k \Gamma(k+1/2)}{\sqrt{\pi}}  (\Sigma_1(t))^k \,. 
\end{equation}
From the latter equation we obtain the typical value of $S_{1}(\omega=0,t)$:
\begin{equation}
\label{nn2}
S_1^\mathrm{typ}(\omega=0,t) = \exp\left(\overline{\ln S_1(\omega=0,t)}\right) 
= \frac{1}{2} e^{-\gamma_{\mathrm{E}}}\,\Sigma_1(t) \,,
\end{equation}
where $\gamma_{\mathrm{E}}$ is the Euler gamma constant. Equations \eqref{P3} to~\eqref{nn2} allow us to draw the following conclusions:
\begin{itemize}
\item  Similarly to the case $t \to \infty$ and $\omega > 0$, 
the entire dependence of the power spectral density $S_1(\omega=0,t)$ on $t$, (and similarly, of $S_2(\omega=0,t)$ on $t$), for a given realization of noises, 
is encoded in the ensemble-averaged value $\Sigma_1(t)=\overline{S_1(\omega=0,t)}$ (or $\Sigma_2(t)=\overline{S_2(\omega=0,t)}$). This implies that Eq.~\eqref{P3} can be rewritten as an equality in distribution:
\begin{align}
\label{rr}
\frac{S_1(\omega=0,t)}{\Sigma_1(t)} \overset{\mathrm{d}}{=} r \,,
\end{align}
where $r$ is a random dimensionless variable with a universal distribution $P(r) = \exp(- r/2)/\sqrt{2 \pi r}$. 
Therefore, the form of $\Sigma_1(t)$ can also be inferred from \textit{a single realization} of $V_1$, without resorting to ensemble averaging.\\
\item The probability density function in Eq. \eqref{P3} is effectively broader than the one in Eq.~\eqref{distS}. Indeed, one finds that the standard deviation is given by
 \begin{align}
\sigma_1(\omega=0,t) = \sqrt{2} \,  \Sigma_1(t)\,.
 \end{align}
Therefore the fluctuations around the mean value are \textit{larger} than the mean, so
that the coefficient of variation $\mathsf{cv}(t) = \sigma_1(\omega=0,t)/\Sigma_1(t)$ of the distribution in Eq.~\eqref{P3}  is exactly equal to $\sqrt{2}$. 
This explains, in fact, why we do observe a bigger scatter of the data in the vicinity of $\omega = 0$ in Fig. \ref{fig2}, than for $\omega > 0$. \\
\item  $S_1^\mathrm{typ}(\omega=0,t)$ in Eq.~\eqref{nn2}  depends on $t$ exactly
as its ensemble-averaged counterpart does, and has a smaller amplitude (compare with~\eqref{typ}). 
This suggests again that the ensemble-averaged property $\Sigma_1(t)=\overline{S_1(\omega=0,t)}$ is supported by some \textit{atypical} realizations of $V_1$.  
\end{itemize}

\subsection{Probability distribution of~$\chi_t$ (Eq.~\eqref{chi})}

We now dwell on the observable~$\chi_t$ defined in Eq.~\eqref{chi} for \textit{given} trajectories $V_1(\tau)$ and $V_2(\tau)$. This quantity is a random variable with the dimensions of temperature.  Its first moment is equal to $\overline{\chi}_t$, given in Eq.~\eqref{parchi} in the Appendix C (and hence, in the limit $t \to \infty$, to $\overline{\chi}$ in Eq.~\eqref{Rparameter}). In what follows, we evaluate its probability distribution function $P(\chi_t)$, which encodes all the information how $\chi_t$ varies from one realization of noises to another.

Such a probability density function can be calculated in a standard way by introducing first the moment-generating function, and then performing a corresponding inverse Fourier transform (see Appendix \ref{C} for the details of derivations). In doing so, we find that in the limit $t \to \infty$, the limiting form 
of $P(\chi_t)$ is given explicitly by
\begin{equation}
\label{distchi}
P(\chi_t) = \frac{1}{8 \pi \sqrt{T_1 T_2}} \exp\left(\frac{T_1 - T_2}{16 T_1 T_2} \chi_t \right) K_0\left(\frac{T_1+ T_2}{16  T_1 T_2} |\chi_t|\right) \,, 
\end{equation}
where $K_0(x)$ is  the modified Bessel function  of the second kind. The distribution $P(\chi_t)$ for an arbitrary $t$ is reported in Appendix~\ref{C}. 

Note that the expression~\eqref{distchi} is sharply peaked at $\chi_t = 0$ (see also Fig. \ref{fig5}). In fact, $P(\chi_t)$ diverges logarithmically when $\chi_t \to 0$.  This implies that typically (i.e., for most of realizations of the noises) one will observe small values of the parameter $\chi_t$, and therefore of its average, regardless of whether the dynamics proceeds in equilibrium or in non-equilibrium. Hence, the average of the random variable $\chi_t$ does not permit us  by itself to distinguish between equilibrium and non-equilibrium dynamics. However, from the comparison of the result in Eq. \eqref{distchi} against the experimental data,  we see that its distribution can be obtained with comparatively small statistics. Now, the shape of $P(\chi_t)$ allows us to discriminate between equilibrium and non-equilibrium. Indeed, $P(\chi_t)$ is symmetric  around the origin and narrow in equilibrium, whereas it is \textit{asymmetric} and substantially broader out of equilibrium.

We now consider the tails of the distribution in  Eq.~\eqref{distchi}. In the limit $t\to\infty$, we have for large positive values of~$\chi_t$
\begin{align}
P\left(\chi_t\right) \simeq \frac{1}{\sqrt{8 \pi (T_1 + T_2) \chi_t}} \exp\left( - \frac{\chi_t}{8 T_1}\right) \,,
\end{align}
while for $\chi_t \to - \infty$ we obtain
\begin{align}
P\left(\chi_t\right) \simeq \frac{1}{\sqrt{8 \pi (T_1 + T_2) |\chi_t|}} \exp\left( \frac{\chi_t}{8 T_2}\right) \,,
\end{align}
So, the decay is a bit faster than exponential (due to an additional power law) and remarkably, the right tail (large positive values of $\chi_t$) is entirely controlled by $T_1$, while the left (large negative values of $\chi_t$) one is entirely controlled by $T_2$.

\section{Concluding remarks}
\label{conc}
In this paper we have considered a Brownian gyrator model, described by two coupled oscillators,  connected to two thermal reservoirs kept at different temperatures,  $T_1$ and $T_2$. Models of this kind are commonly used to describe non-equilibrium and fluctuating mesoscopic devices, as well as are of interest in current nano- and bio-technologies. One important question we have addressed here is whether one can conclusively distinguish between the evolution in 
non-equilibrium conditions ($T_1 \ne T_2$), when the system performs rotations around the origin, and in equilibrium ones ($T_1 = T_2$) 
when no such a rotation takes place, considering not the trajectories of the system, but rather their spectral densities. 
We have provided two relatively simple criteria to ascertain whether the system evolves under equilibrium or non-equilibrium conditions. 

First, Eqs.~\eqref{zz} show that the maxima of the power spectral densities of the voltages at zero frequency, $\mu_1(0)$ and~$\mu_2(0)$ depend linearly on the resistances $R_i$ and on the temperature of the respective baths $T_i$, $i=1,2$. Therefore, knowledge of the resistances and measurements of the power spectra will reveal whether the system is or is not in equilibrium. Moreover, whenever $R_1=R_2$, their value does not need to be known: $\mu_1(0) = \mu_2(0)$ means equilibrium state, while $\mu_1(0) \ne \mu_2(0)$ means non-equilibrium state, where the spectral densities are computed along a single realization of the process.

Second, we have shown that in the small $\omega$ regime, the Pearson coefficient for the power spectral densities
constructed from a single realization of the process scales differently with the frequency $\omega$ in equilibrium and in non-equilibrium conditions. Indeed, $\rho_{\omega} \sim \omega^4$ for $T_1 = T_2$, while $\rho_{\omega} \sim \omega^2$ for the $T_1 \ne T_2$, cf.~Eq.~\eqref{asymp}. For such a criterion no knowledge of the material parameters is required.

These expressions can in principle be tested experimentally in systems such as those reported in Refs~\cite{al1,al2}, and more recently in Ref.~\cite{CilMD}. The last paper, in particular, illustrates the realization of a Maxwell demon that reverses the direction of heat flux between two electric circuits kept at different temperatures and coupled by the thermal noise. Therefore applying the tests here proposed to such a system is quite intriguing, due to the peculiar behavior of this device.  In general, our approach based on the spectral properties of individual trajectories of systems evolving in and out of equilibrium opens new perspectives for the analysis of non-equilibrium phenomena.

\section*{Acknowledgments}
The authors wish to thank A. Puglisi and A. Vulpiani for insightful discussions and some useful suggestions. 
Research of L.R. is supported by the "Departments of Excellence 2018-2022" grant awarded by the Italian Ministry of 
Education, University and Research (MIUR), grant E11G18000350001.



\begin{widetext}
\appendix
\section{Solution of eqs.~\eqref{langevin_currents}}
\label{sol1}
Let $\tilde{q}_{i}$ and $\tilde{\eta}_{i}$ denote the Laplace-transformed charges and noises, 
\begin{align}
\tilde{q}_{i} = {\cal L}[q_{i}] = \int^{\infty}_0 d\tau \exp(-\lambda \tau) \, q_{i} \,,
\end{align}
and
\begin{align}
\tilde{\eta}_{i} = {\cal L}[\eta_{i}] = \int^{\infty}_0 d\tau \exp(-\lambda \tau) \, \eta_{i} \,,
\end{align}
respectively. Then, the Laplace-transformed system of Eqs.\eqref{langevin_currents} reads:
\begin{equation}
\begin{split}
\lambda R_1 \, \tilde{q}_1 &= - \frac{C_2}{X} \tilde{q}_1 + \frac{C}{X} \left(\tilde{q}_2 - \tilde{q}_1\right) + \tilde{\eta}_1 \,,  \\
\lambda R_2 \, \tilde{q}_2  &= - \frac{C_1}{X} \tilde{q}_2 + \frac{C}{X} \left(\tilde{q}_1 - \tilde{q}_2\right) + \tilde{\eta}_2 \,, 
\end{split}
\end{equation}
and its Laplace-transformed solution is given by
\begin{equation}
\label{solution}
\begin{split}
\tilde{q}_1 &= \frac{\left(C_1 + C + \lambda R_2 X\right) \tilde{\eta}_1 + C \tilde{\eta}_2}{1 +  \left(R_1 (C + C_1) + R_2 (C + C_2)\right) \lambda + R_1 R_2 X \lambda^2} \,, \\
\tilde{q}_2 &= \frac{ C \tilde{\eta}_1 +  \left(C_2 + C + \lambda R_1 X\right) \tilde{\eta}_2}{1 +  \left(R_1 (C + C_1) + R_2 (C + C_2)\right) \lambda + R_1 R_2 X \lambda^2} \,.
\end{split}
\end{equation}
Further, the inverse Laplace transforms ${\cal L}^{-1}$ of the rational functions entering Eqs.~\eqref{solution} are given by
\begin{equation}
\begin{split}
 {\cal L}^{-1}\left[\frac{1}{1 + \left(R_1 (C + C_1) + R_2 (C + C_2)\right) \lambda + R_1 R_2 X \lambda^2}\right] &= \frac{\exp\left(- v \tau\right)}{R_1 R_2 X b}  \sinh\left(b \tau\right) \,, \\
{\cal L}^{-1}\left[ \frac{\lambda}{1 + \left(R_1 (C + C_1) + R_2 (C + C_2)\right) \lambda + R_1 R_2 X \lambda^2}\right]  &=  \frac{\exp\left(- v \tau\right)}{R_1 R_2 X} \left[  \cosh\left(b \tau\right) - \frac{v}{b}  \sinh\left(b \tau\right) \right] \, ,
\end{split}
\end{equation}
where $v$ and $b$ are parameters defined by Eqs.~\eqref{param} in the main text. 
Consequently, the charges for a given realization of noises in the time-domain obey:
\begin{equation}
\label{qs}
\begin{split}
q_1 &= \frac{1}{R_1} \int^{\tau}_0 ds \, Q\left(a,\tau - s\right) \eta_1(s) +  \int^{\tau}_0 ds \, P\left(\tau - s\right) \eta_2(s) \,, \\
q_2 &=  \int^{\tau}_0 ds \, P\left(\tau - s\right) \eta_1(s) +  \frac{1}{R_2} \int^{\tau}_0 ds \, Q\left(-a,\tau - s\right) \eta_2(s) \,,
\end{split}
\end{equation}
where $Q$ is defined by Eq.~\eqref{Q}, the parameter $a$ is given by 
\begin{align}
a = \frac{R_2 \left(C + C_2\right)- R_1 \left(C + C_1\right)}{2  R_1 R_2 X} \,,
\end{align}
and the function $P(\tau)$ obeys
\begin{align}
P(\tau) &= \frac{C}{R_1 R_2 X b} \exp\left(- v \tau\right) \sinh\left(b \tau\right)  \,.
\end{align}
In turn, the charges and the measured voltages are related by \cite{al1,al2}:
\begin{equation}
\label{volts}
\begin{split}
q_1 &= C \left(V_1 - V_2\right) + C_1 V_1 \,, \\ 
q_2 &= C \left(V_1 - V_2\right) - C_2 V_2 \,,
\end{split}
\end{equation}
that, solved for $V_1$ and $V_2$, lead to Eqs.~\eqref{voltages}. 
Substituting Eqs.~\eqref{qs} into Eqs.~\eqref{voltages}, we eventually obtain Eqs.~\eqref{solution}.

\section{Two-time correlation functions of the voltages and the first moment of spectral densities}
\label{B}
The two-time correlation functions $\overline{V_i(\tau_1) V_i(\tau_2)} $, $i=1,2$, are evidently symmetric functions of the time variables $\tau_1$ and $\tau_2$, 
thus it suffice to consider a time-ordered case only. Taking, for instance, $\tau_1 \geq \tau_2 \geq 0$, we get:
\begin{equation}
\label{V1}
\begin{split}
\overline{V_1(\tau_1) V_1(\tau_2)} =  \frac{2 k_\mathrm{B} T_1 (C + C_2)^2}{R_1 X^2} h(a_1) + \frac{2 k_\mathrm{B} T_2 C^2}{R_2 X^2} g \,,
\end{split}
\end{equation}
and
\begin{align}
\label{V2}
\overline{V_2(\tau_1) V_2(\tau_2)} =  \frac{2 k_\mathrm{B} T_1 C^2}{R_1 X^2} g + \frac{2 k_\mathrm{B} T_2 (C+C_1)^2}{R_2 X^2} h(a_2) \,,
\end{align}
where
\begin{equation}
\label{h}
\begin{split}
h(a) = \frac{1}{4 v b^2 (v^2 - b^2)} & e^{-v (\tau_1 + \tau_2)} \Bigg(
\Bigg[
 b^2 v^2 \left(2 e^{2 v \tau_2} -1 \right)
+ a^2 \left(b^2 \left(e^{2 v \tau_2} - 1\right) + v^2\right)  -  b^4 \left(e^{2 v \tau_2} - 1\right) - 2 a  v b^2 \, e^{2 v \tau_2}\Bigg] \\
& \times  \cosh\left(b (\tau_1 - \tau_2)\right)
+ v \Bigg[\left(2 a b^2 - v (a^2 + b^2)\right) \cosh\left(b (\tau_1 + \tau_2)\right)
+ b \left(a^2 + b^2 - 2 a v \right) \\
&\times \Big(e^{2 v \tau_2} \sinh\left(b (\tau_1 - \tau_2)\right) - \sinh\left(b (\tau_1 + \tau_2)\right)
\Big)
\Bigg]
\Bigg) \,,
\end{split}
\end{equation}
and
\begin{equation}
\label{g}
\begin{split}
g = \frac{1}{4 v b^2} e^{- v (\tau_1 + \tau_2)} &\Bigg(\Big(b^2 \left(e^{2 v \tau_2} - 1\right)+ v^2\Big) \cosh\left(b (\tau_1 - \tau_2)\right) - v \Big[ v \cosh\left(b (\tau_1 + \tau_2)\right) + \\ 
&+b \, e^{2 v \tau_2} \sinh\left(b (\tau_1 - \tau_2)\right) - b \sinh\left(b (\tau_1 + \tau_2)\right) \Big]
\Bigg) \,.
\end{split}
\end{equation}
The dependence of both $h(a)$ and $g$ not only on the difference $\tau_1 - \tau_2$ but also on the sum $\tau_1 + \tau_2$, indicates that the initial condition $V_1(\tau=0) = V_2(\tau=0) = 0$ needs to evolve in time, in order to approach a stationary state. In fact, the $\tau_1 \to \infty$ limit, with fixed $\tau_2$, makes
the two voltages decouple, and both $h(a)$ and $g$ vanish. On the other hand, recalling that $v > b$, cf.\ Eq.\eqref{param}, 
taking $\tau_1 = \tau_2 = \tau$, and letting $\tau$ grow without bounds one obtains:
\begin{equation}
\begin{split}
\lim_{\tau \to \infty} h(a) &=  \frac{1}{4 v} + \frac{(v - a)^2}{4 v (v^2 - b^2)} \,, \\
\lim_{\tau \to \infty} g &=  \frac{1}{4 v}\,, \nonumber\\
\end{split}
\end{equation}
Consequently, in this limit the variances of the measured voltages converge as follows:
\begin{equation}
\label{Vs}
\begin{split}
\lim_{\tau \to \infty} {\rm Var}\left(V_1(\tau)\right) &= \lim_{\tau \to \infty}  \overline{V_1(\tau) V_1(\tau)} = \frac{ (C + C_2)^2}{2 R_1 X^2 v} \left(1 + \frac{(v - a_1)^2}{v^2 - b^2} \right)  k_\mathrm{B} T_1 + \frac{ C^2}{2 R_2 X^2 v}  k_\mathrm{B} T_2\,, \\
\lim_{\tau \to \infty} {\rm Var}\left(V_2(\tau)\right) &= \lim_{\tau \to \infty}  \overline{V_2(\tau) V_2(\tau)} =  \frac{ C^2}{2 R_1 X^2 v}  k_\mathrm{B} T_1 + \frac{ (C+C_1)^2}{2 R_2 X^2 v} \left(1 + \frac{(v - a_2)^2}{v^2 - b^2} \right)  k_\mathrm{B} T_2\,.
\end{split}
\end{equation}
These variances saturate at finite values, when time grows, consistently with the fact that we have two coupled Ornstein-Uhlenbeck processes, cf.~Eqs. \eqref{langevin_currents}. Explicit forms of their limiting values, in terms of the physical parameters can be found by merely substituting the 
definitions of  $v$, $b$, $a_1$ and $a_2$ introduced in Eq.~\eqref{param}. 

Concerning the cross-correlation function of the voltages, with $\tau_1 \geq \tau_2 \geq 0$, we get:
\begin{equation}
\label{V12}
\begin{split}
\overline{V_1(\tau_1) V_2(\tau_2)} &= \frac{k_\mathrm{B} T_1 C (C+C_2)}{2 R_1 X^2 b^2 v} \exp\left(- v (\tau_1 + \tau_2)\right) \Bigg(\left(b^2 \left(e^{2 v \tau_2} - 1\right) + v a_1\right) \cosh\left(b (\tau_1 - \tau_2)\right) \\
&\qquad {}- a_1 v \cosh\left(b (\tau_1 + \tau_2)\right) - b \left(a_1 \left(e^{2 v \tau_2} - 1\right) + v\right) \sinh\left(b (\tau_1 - \tau_2)\right) + b v \sinh\left(b (\tau_1 + \tau_2)\right)
\Bigg) \\
&{}+ \frac{k_\mathrm{B} T_2 C (C+C_1)}{2 R_2 X^2 b^2 v} \exp\left(- v (\tau_1 + \tau_2)\right)  \Bigg(\left(b^2 \left(e^{2 v \tau_2} - 1\right) + v a_2\right) \cosh\left(b (\tau_1 - \tau_2)\right) \\
&\qquad {}- a_2 v \cosh\left(b (\tau_1 + \tau_2)\right) + b \left(a_2 \left(e^{2 v \tau_2} -1\right)\right.\\
&\qquad \left. {} + v \left(1 - 2 e^{2 v \tau_1}\right)  \right)
\sinh\left(b (\tau_1 - \tau_2)\right) + b v \sinh\left(b (\tau_1 + \tau_2)\right)\Bigg) \,.
\end{split}
\end{equation}
which vanishes when $\tau_1 = \tau_2 = 0$, and also when $\tau_1 \to \infty$ with fixed $\tau_2$, which signifies that correlations between $V_1$ and $V_2$ 
decouple in this limit. Next, for $\tau_1 = \tau_2 = \tau$ and $\tau \to \infty$, the expression \eqref{V12} attains a constant value
\begin{align}
\label{V12lim}
\lim_{\tau \to \infty}\overline{V_1(\tau) V_2(\tau)} &= \frac{ C (C+C_2)}{2 R_1 X^2  v} k_\mathrm{B} T_1 + \frac{ C (C+C_1)}{2 R_2 X^2  v} k_\mathrm{B} T_2 \,.
\end{align} 
Note now that the limiting values of the variances of the measured voltages are monotonically increasing functions of the temperatures $T_1$ and $T_2$, as they should, and so is the limiting value of the cross-correlations, Eqs. \eqref{V12lim}.   
Moreover, expressions \eqref{Vs} and \eqref{V12lim} do not undergo any qualitative change with respect to the 
non-equilibrium case, when $T_1$ turns equal to $T_2$.  Further, the time scales that characterize the
relaxation to the stationary state on the material parameters $v$ and $b$ only, and are independent of the temperatures. 
In conclusion, because we deal here with Gaussian stochastic processes, 
which are fully  defined by the first two moments, it seems unlikely that a useful criterion for distinguishing  
non-equilibrium and equilibrium behaviors be based solely on the correlation functions of voltages. 

Equations \eqref{V1} to \eqref{g} straightforwardly lead to the expression of the power spectral densities of the processes $V_1(\tau)$ and $V_2(\tau)$, which are first moments of the random variables $S_1$ and $S_2$, Eqs. \eqref{spectra}, in the limit of long observation time $t$. Let us then consider the following quantities:
\begin{equation}
\label{hfgf}
\begin{split}
h_{\omega}(a) &= \lim_{t \to \infty} \frac{1}{t} \int^t_0 d\tau_1  \int^t_0 d\tau_2 \cos\left(\omega (\tau_1 - \tau_2)\right) h(a) = 2  \lim_{t \to \infty} \frac{1}{t} \int^t_0 d\tau_1  \int^{\tau_1}_0 d\tau_2 \cos\left(\omega (\tau_1 - \tau_2)\right) h(a) \,, \\
g_{\omega} &= \lim_{t \to \infty} \frac{1}{t} \int^t_0 d\tau_1  \int^t_0 d\tau_2 \cos\left(\omega (\tau_1 - \tau_2)\right) g =  2 \lim_{t \to \infty} \frac{1}{t} \int^t_0 d\tau_1  \int^{\tau_1}_0 d\tau_2 \cos\left(\omega (\tau_1 - \tau_2)\right) g \,.
\end{split}
\end{equation}
Substituting Eqs.\eqref{h} and \eqref{g} in Eq. \eqref{hfgf}, integrating over the time variables, and taking the $t \to \infty$ limit, yields: 
\begin{equation}
\label{forms}
\begin{split}
h_{\omega}(a) &= \frac{2 (\omega^2 + \left(v-a\right)^2)}{b^4 + 2 b^2 \left(\omega^2 - v^2\right) + \left(\omega^2 + v^2\right)^2} \,,\\
g_{\omega} &=  \frac{2 \omega^2}{b^4 + 2 b^2 \left(\omega^2 - v^2\right) + \left(\omega^2 + v^2\right)^2} \,.
\end{split}
\end{equation}
which together with Eqs.~\eqref{V1} and \eqref{V2} leads to Eqs.~\eqref{eq:mean_spectra}.   

\section{Bivariate moment-generating and probability density functions of single-trajectory spectral densities}
\label{C}
Let us rewrite the definitions of the single-realization spectral densities, Eq.\eqref{spectra}, in the following form :
\begin{equation}
\label{spectra}
\begin{split}
S_1 &=  \frac{1}{t} \left(\int^t_0 d\tau \cos\left(\omega \tau\right) V_1(\tau)\right)^2 + \frac{1}{t} \left(\int^t_0 d\tau \sin\left(\omega \tau\right) V_1(\tau)\right)^2\,, \\
S_2 &= \frac{1}{t} \left(\int^t_0 d\tau \cos\left(\omega \tau\right) V_2(\tau)\right)^2 + \frac{1}{t} \left(\int^t_0 d\tau \sin\left(\omega \tau\right) V_2(\tau)\right)^2\,. 
\end{split}
\end{equation}
Then, using the integral identity
\begin{align}
e^{-B^2/4 A} = \sqrt{\frac{A}{\pi}} \int^{\infty}_{-\infty} dz \, e^{- A z^2 + i B z} \,,
\end{align}
the bivariate moment-generating function, Eq. \eqref{Phi}, can be written as a four-fold integral:
\begin{equation}
\label{bva}
\begin{split}
\Phi(\lambda_1,\lambda_2) = \frac{1}{\left(4 \pi\right)^2 \lambda_1 \lambda_2}\int^{\infty}_{-\infty} dz_1 \int^{\infty}_{-\infty} dz_2 \int^{\infty}_{-\infty} dz_3 \int^{\infty}_{-\infty} dz_4 \, \exp\left(- \frac{z_1^2 + z_2^2}{4 \lambda_1} - \frac{z_3^2 + z_4^2}{4 \lambda_2}\right) {\cal R}({\bf z}) \,,
\end{split}
\end{equation}
in which the function ${\cal R}({\bf z})$ is given by
\begin{equation}
\label{R}
\begin{split}
{\cal R}({\bf z}) &= \overline{\exp\left(\frac{i}{\sqrt{t}} \int^t_0 d\tau \int^{\tau}_0 ds \, F_{\omega}(\tau,s) \eta_1(s)\right)}     \\ &\times \overline{\exp\left( - \frac{i}{\sqrt{t}} \int^t_0 d\tau \int^{\tau}_0 ds \, G_{\omega}(\tau,s) \eta_2(s)  \right)} \\
& = \exp\left(- \frac{2  k_\mathrm{B}  T_1 R_1}{t} \int^t_0 d\tau_1 \int^t_0 d\tau_2 \int^{{\rm min}(\tau_1,\tau_2)}_0 ds \, F_{\omega}(\tau_1,s) \, F_{\omega}(\tau_2,s)\right) \\
& \times \exp\left(- \frac{2  k_\mathrm{B}  T_2 R_2}{t} \int^t_0 d\tau_1 \int^t_0 d\tau_2 \int^{{\rm min}(\tau_1,\tau_2)}_0 ds \, G_{\omega}(\tau_1,s) \, G_{\omega}(\tau_2,s) \right)\,,
\end{split}
\end{equation}
where the bar in the first line stands for averaging over the Gaussian white-noise $\eta_1$, the bar in the second line denotes averaging over 
$\eta_2$, which is statistically independent of $\eta_1$, while the functions $F_{\omega}(\tau,s)$ and $G_{\omega}(\tau,s)$ are expressed by:
\begin{equation}
\label{FG}
\begin{split}
F_{\omega}(\tau,s) &= \frac{C+C_2}{R_1 X} \Big(z_1 \cos(\omega \tau) + z_2 \sin(\omega \tau)\Big) Q(a_1, \tau-s) + \frac{C}{R_1 X} \Big(z_3 \cos(\omega \tau) + z_4 \sin(\omega \tau)\Big) Q(v,\tau-s) \,, \\
G_{\omega}(\tau,s)&= \frac{C}{R_2 X} \Big(z_1 \cos(\omega \tau) + z_2 \sin(\omega \tau)\Big) Q(v, \tau-s) + \frac{C+C_1}{R_2 X}  \Big(z_3 \cos(\omega \tau) + z_4 \sin(\omega \tau)\Big) Q(a_2,\tau-s) \,,
\end{split}
\end{equation}
where $Q(a_1, \tau-s)$, $Q(v,\tau-s)$ and $Q(a_2,\tau-s)$ are defined in the main text in Eq. \eqref{Q} (with $a$ replaced by $a_1$, $v$ and $a_2$, respectively).

Introducing the Heaviside theta-function, defined by  $\theta(\tau) = 1$ for $\tau \geq 0$, and $\theta(\tau) = 0$ for $\tau < 0$, we can write:
\begin{equation}
\label{F}
\begin{split}
&\int^t_0 d\tau_1 \int^t_0 d\tau_2 \int^{{\rm min}(\tau_1,\tau_2)}_0 ds \, F_{\omega}(\tau_1,s) \, F_{\omega}(\tau_2,s) = \\
&=  \int^t_0 d\tau_1 \int^t_0 d\tau_2 \int^{t}_0 ds \, F_{\omega}(\tau_1,s) \, F_{\omega}(\tau_2,s) \,\theta(\tau_2 - s) \,\theta(\tau_1 - s)  =    \int^t_0 ds \left(\int^t_s d\tau \, F_{\omega}(\tau,s) \right)^2\,,  \\
\end{split}
\end{equation}
and 
\begin{equation}
\label{G}
\begin{split}
&\int^t_0 d\tau_1 \int^t_0 d\tau_2 \int^{{\rm min}(\tau_1,\tau_2)}_0 ds \, G_{\omega}(\tau_1,s) \, G_{\omega}(\tau_2,s)  = \\
& \int^t_0 d\tau_1 \int^t_0 d\tau_2 \int^{t}_0 ds \, G_{\omega}(\tau_1,s) \, G_{\omega}(\tau_2,s) \,\theta(\tau_2 - s) \, \theta(\tau_1 - s) =   \int^t_0 ds \left(\int^t_s d\tau \, G_{\omega}(\tau,s) \right)^2 \,.  \\
\end{split}
\end{equation}
The above expressions are fairly general and our subsequent analysis focuses on two particular limits: A standard text-book limit of an infinitely large observation time $t$, and the limit $\omega \equiv 0$ and arbitrary $t$.

\subsection{Limit $t \to \infty$ and $\omega > 0$.}

In the limit $t \to \infty$, rather cumbersome calculations eventually lead to the following expressions:
\begin{equation}
\begin{split}
&\lim_{t \to \infty} \frac{1}{t} \int^t_0 ds \left(\int^t_s d\tau \, F_{\omega}(\tau,s) \right)^2 = \\
&\qquad = \frac{\left(C+C_2\right)^2 \left(\omega^2 + \left(v - a_1\right)^2\right)}{2 R_1^2 X^2 \left(b^4 + 2 b^2 (\omega^2 - v^2) + \left(\omega^2 + v^2\right)^2\right)} \left(z_1^2 + z_2^2\right)\\
&\qquad\qquad + \frac{C^2 \omega^2}{2 R_1^2 X^2 \left(b^4 + 2 b^2 (\omega^2 - v^2) + \left(\omega^2 + v^2\right)^2\right)}  \left(z_3^2 + z_4^2\right) \\
&\qquad\qquad {}+\frac{C \left(C+C_2\right) \, \omega \, \Big(\omega \left(z_1 z_3 + z_2 z_4\right) + (v-a_1) \left(z_2 z_3 - z_1 z_4\right) \Big)}{R_1^2 X^2 \left(b^4 + 2 b^2 (\omega^2 - v^2) + \left(\omega^2 + v^2\right)^2\right)} \,, \\
&\lim_{t \to \infty} \frac{1}{t} \int^t_0 ds \left(\int^t_s d\tau \, G_{\omega}(\tau,s) \right)^2 = \\
&\qquad =\frac{C^2 \omega^2}{2 R_2^2 X^2 \left(b^4 + 2 b^2 (\omega^2 - v^2) + \left(\omega^2 + v^2\right)^2\right)}  \left(z_1^2 + z_2^2\right)\\
&\qquad\qquad +
\frac{\left(C+C_1\right)^2 \left(\omega^2 + \left(v - a_2\right)^2\right)}{2 R_2^2 X^2 \left(b^4 + 2 b^2 (\omega^2 - v^2) + \left(\omega^2 + v^2\right)^2\right)} \left(z_3^2 + z_4^2\right)   \\
&\qquad\qquad + \frac{C (C + C_1) \omega \Big(\omega \left(z_1 z_3 + z_2 z_4\right) - (v-a_2) \left(z_2 z_3 - z_1 z_4\right) \Big)}{R_2^2 X^2 \left(b^4 + 2 b^2 (\omega^2 - v^2) + \left(\omega^2 + v^2\right)^2\right)} \,.
\end{split}
\end{equation}
Note that the leading subdominant time-dependent terms in the integrals in Eqs. \eqref{F} and \eqref{G} are of order of $O(1/t)$, with amplitudes containing $\sin(\omega t)$ 
and $\cos(\omega t)$), while the remaining terms decay exponentially fast: $\exp(- 2 (v - b) t)$.

Recalling next the explicit expressions for the first moments of random variables $S_1$ and $S_2$, Eqs. \eqref{eq:mean_spectra},
we find that the function ${\cal R}({\bf z})$ in Eq. \eqref{R} attains the following form
\begin{equation}
\label{RR}
{\cal R}({\bf z}) = \exp\Bigg(- \frac{\mu_1(\omega)}{4} \left(z_1^2 + z_2^2\right) - \frac{\mu_2(\omega)}{4} \left(z_3^2 + z_4^2\right) - A  \left(z_1 z_3 + z_2 z_4\right) - 
B \left(z_2 z_3 - z_1 z_4\right)\Bigg) \, ,
\end{equation}
with 
\begin{align}
A = \frac{2 k_\mathrm{B} C \omega^2}{X^2 \left(b^4 + 2 b^2 (\omega^2 - v^2) + \left(\omega^2 + v^2\right)^2\right)} \left(\frac{(C+C_2)}{R_1} T_1 +  \frac{(C+C_1)}{R_2} T_2\right) \,,
\end{align}
and 
\begin{align}
B= \frac{2 k_\mathrm{B} C \omega}{X^2 \left(b^4 + 2 b^2 (\omega^2 - v^2) + \left(\omega^2 + v^2\right)^2\right)}  \left(\frac{(C+C_2)(v - a_1)}{R_1} T_1 - \frac{(C+C_1) (v - a_2)}{R_2} T_2\right) \,.
\end{align}
Then, using Eqs.\eqref{RR} and \eqref{bva}, and performing the integrations, we get our expression in Eq. \eqref{mg} with
\begin{align}
\rho_{\omega} = \frac{4 (A^2 + B^2)}{\mu_1(\omega) \mu_2(\omega)} \, .
\end{align}
To prove that $\rho_{\omega}$ is indeed the Pearson's correlation coefficient in  Eq. \eqref{pearson}, we first differentiate the expression in Eq. \eqref{mg} 
with respect to $\lambda_1$ and $\lambda_2$, to get
\begin{align}
\left. \left(\frac{d^2 }{d\lambda_1 d\lambda_2} \Phi(\lambda_1,\lambda_2)\right) \right|_{\lambda_{1,2}=0} = \overline{S_1 S_2} = \mu_1(\omega) \mu_2(\omega) \left(1 + \rho_{\omega}\right) \,,
\end{align}
from which it follows that 
\begin{align}
\rho_{\omega} = \frac{\overline{S_1 S_2} - \mu_1(\omega) \mu_2(\omega)}{\mu_1(\omega) \mu_2(\omega)} =  \frac{\overline{\left(S_1 - \mu_1(\omega)\right) \left( S_2 - \mu_2(\omega)\right)}}{\mu_1(\omega) \mu_2(\omega)} \,.
\end{align}
The proof is completed by merely considering that 
\begin{equation}
\begin{split}
\sqrt{\overline{S_1^2} - \mu_1^2(\omega)} &=\mu_1(\omega) \,, \\
\sqrt{\overline{S_2^2} - \mu_2^2(\omega)} &= \mu_2(\omega) \,.
\end{split}
\end{equation}
for the model under investigation,  i.e., as we have already remarked in the main text, in the limit $t \to \infty$ the standard deviations of the single-trajectory power spectral densities are exactly equal to the corresponding mean values.

Lastly, we invert the expression in Eq. \eqref{mg} with respect to $\lambda_1$ and $\lambda_2$, which is conveniently done by first expanding it into the Taylor series in powers of $\rho_{\omega}$. In doing so, we obtain a series with coefficients that are factorized with respect to the Laplace parameters. Inversion is then straightforward and yields  
\begin{align}
\label{joint20}
P\left(S_1 = s_1,S_2 = s_2\right) = \dfrac{\exp\left(- \dfrac{s_1}{\mu_1(\omega)} - \dfrac{s_2}{\mu_2(\omega)}\right) }{\mu_1(\omega) \mu_2(\omega)}  \sum_{k=0}^{\infty} \rho_{\omega}^k L_k\left(\frac{s_1}{\mu_1(\omega)}\right) L_k\left(\frac{s_2}{\mu_2(\omega)}\right)   \,,
\end{align}
where $L_k$ are the Laguerre polynomials. Summing the series, we arrive at equation~\eqref{joint2}. 


\subsection{Areas under random curves for arbitrary $t$}

We turn next to the particular case $\omega = 0$, i.e., we focus on the statistical properties of integrated voltages $V_1$ and $V_2$. There are two reasons for such an analysis. First, we realized that in the limit $t \to \infty$ the squared areas under random curves $V_1$ and $V_2$ statistically decouple from each other because the Pearson's coefficient vanishes when $\omega =0$, which makes these 
properties important for distinguishing equilibrium and non-equilibrium dynamics. 
The $t \to \infty$ limit, however, should be taken with caution, since no experiment nor numerical simulations 
last an infinite time. Therefore, one needs to know the rate at which corrections to the limiting behavior vanish.

The second reason is a bit more subtle.  
It is known that (see Refs. \cite{g1,g2,vit1,vit2}) for any one-dimensional Gaussian process $X_t$
its moment-generating function, and hence, the ensuing probability density 
of a single-trajectory power-spectral density is entirely defined by the first two moments - the mean and the variance, 
likewise the parental process $X_t$ itself. In general, for finite $\omega$ and $t$, bounded away from zero and infinity, the moment-generating function  and 
the probability density function of a (single-component) single-trajectory power spectral density of a Gaussian process are given by
\begin{align}
\Phi(\lambda_1) &= \frac{1}{\sqrt{1 + 2 \overline{S_1} \lambda_1 + \left(2 - \mathsf{cv}^2\right)  \overline{S_1}^2 \lambda_1^2 }} \label{Ccc} \\
P(S_1 = s_1) &= \frac{1}{\sqrt{2 - \mathsf{cv}^2} \, \overline{S_1}} \exp\left( - \frac{1}{2-\mathsf{cv}^2} 
\frac{s_1}{\overline{S_1} }\right) I_0\left(\frac{\sqrt{\mathsf{cv}^2 - 1}}{2-\mathsf{cv}^2} \frac{s_1}{\overline{S_1} }\right) \label{gaussian}
\end{align}
where $\mathsf{cv} = \sigma_1(\omega,t)/\overline{S_1}$ is the coefficient of variation, which obeys  $1 \leq \mathsf{cv}^2 \leq 2$. 
The probability density function in Eq. \eqref{gaussian},
the so-called Bessel function distribution, attains, for a sub-diffusive Gaussian processes $X_t$, a simpler form
in the limit when either $t$ or $\omega$ tend to infinity, as in Eq. \eqref{distS}, since $\mathsf{cv}$ approaches unity in this limit.  
In this regard, Eq. \eqref{distS} is completely in line with the general results of Refs. \cite{g1,g2}. On the other hand, 
when the limit $t \to \infty$ is taken first, a subsequent passage to the limit $\omega \to 0$ yields a spurious behavior. The point is 
that the first moment of $S_1$ 
and the standard deviation $\sigma_1$ from the mean value are functions of both $\omega$ and $t$, where the observation time enters 
as a product $\omega t$ \cite{g1,g2,vit1,vit2}. Therefore, taking the limit  $t \to \infty$ first we naturally get only a part of the actual 
dependence on $\omega$ and, as a consequence, an incorrect limiting behavior. As a matter fact, for $\omega=0$ and finite $t$, the 
coefficient of variation obeys $\mathsf{cv} = \sqrt{2}$, so that the 
Bessel function distribution reduces to a simpler form, but its functional form is different from that in 
Eq. \eqref{distS}. This prompts us to focus on the limit $\omega = 0$ at finite $t$ for the model under study 
involving two coupled components $V_1$ and $V_2$. 

Our starting point is the general expression in our Eq. \eqref{bva}. Setting $\omega = 0$ in Eqs.~\eqref{FG} we have that the latter attain the form
\begin{equation}
\label{FG0}
\begin{split}
F_0(\tau,s) &= \frac{C+C_2}{R_1 X} z_1  Q(a_1, \tau-s) + \frac{C}{R_1 X} z_3 Q(v,\tau-s) \,, \\
G_0(\tau,s)&= \frac{C}{R_2 X} z_1  Q(v, \tau-s) + \frac{C+C_1}{R_2 X} z_3  Q(a_2,\tau-s) \,,
\end{split}
\end{equation}
which no longer depend on $z_2$ and $z_4$. Inserting these expressions into Eq. \eqref{bva} and integrating, we find
our Eq. \eqref{Phif0} presented in main text, 
where $\Sigma_1(t)=\overline{S_1(\omega=0,t)}$ and $\Sigma_2(t)=\overline{S_2(\omega=0,t)}$ are averaged spectral densities of $V_1$ and $V_2$ calculated at zero frequency, 
i.e., are the averaged squared areas below $V_1$ and $V_2$ divided by $t$, while $\rho_0$ corresponds to the $\omega=0$ Pearson coefficient 
\begin{align}
\label{pearson4}
\rho_0 =  
\frac{\overline{
\left(S_1(\omega=0,t) - \Sigma_1(t)\right) \left(S_2(\omega=0,t) - \Sigma_2(t)\right)}}{\sqrt{\left(\overline{S_1^2(\omega=0,t)} - \Sigma_1(t)^2\right) \left(\overline{S_2^2(\omega=0,t)} - \Sigma_2(t)^2\right)}}  \,.
\end{align} 
For arbitrary $t$, the exact expressions for $\Sigma_1(t)$, $\Sigma_2(t)$ and $\rho_0$ have quite a complicated form which we do not present here. Overall, $\Sigma_1(t)$ and $\Sigma_2(t)$ are monotonically increasing functions of $t$ (see solid curves in Fig. \ref{figA3}, panels (a) and (b)), which are characterized by an initial  power-law growth stage and ultimately, by  
a relaxation towards their asymptotic values. For small values of $t$ we have the parabolic growth laws of the form :
\begin{equation}
\label{Sshortt}
\begin{split}
\Sigma_1(t) &= \frac{4 k_\mathrm{B} \left(C^2 \left(R_2 T_1 + R_1 T_2\right) + C_2 \left(2 C + C_2\right) R_2 T_1\right) }{3 R_1 R_2 X^2} \, t^2 + O\left(t^4\right)  \,, \\
\Sigma_2(t) &=  \frac{4 k_\mathrm{B} \left(C^2 \left(R_2 T_1 + R_1 T_2\right) + C_1 \left(2 C + C_1\right) R_1 T_2\right) }{3 R_1 R_2 X^2} \, t^2 + O\left(t^4\right) \,,
\end{split}
\end{equation}
i.e., $\Sigma_1(t) $ and $\Sigma_2(t)$ vanish at $t = 0$, 
as they should.  The symbol $O\left(t^4\right)$ means that the omitted terms grow with $t$ in proportion to the fourth power of the observations time.

\begin{figure}[htbp]
	\begin{center}
		\includegraphics[width=0.38\hsize]{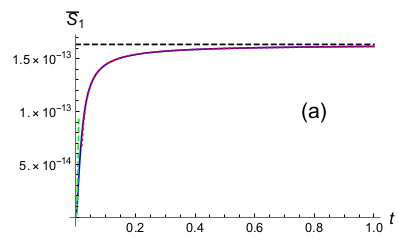}
		\includegraphics[width=0.38\hsize]{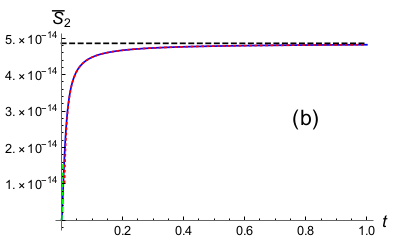}
		\includegraphics[width=0.38\hsize]{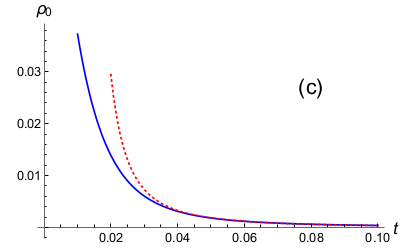}
	\end{center}
	\caption{(Color online) The case $\omega = 0$ with arbitrary $t$. Panel (a): $\overline{S}_1 = \Sigma_1(t)$ ($\unit{V^2s}$) versus the observation time $t$ ($\unit{s}$). Solid (blue) curve represents the full exact expression; green dashed curve depicts the asymptotic form in the first line in Eqs.~\eqref{Sshortt}, while the red dotted curve depicts the large-$t$ asymptotic curve in the first line in Eqs.~\eqref{Slarget}, in which we omit the exponential in $t$ relaxation terms. The horizontal 
(black) dashed line indicates the asymptotic value   $\overline{S}_1 = \overline{S_1(\omega=0,t = \infty)} = 4 R_1 k_\mathrm{B} T_1$.
Panel (b): $\overline{S}_2 = \Sigma_2(t)$ versus the observation time $t$. Solid (blue) curve represents the full exact expression; green dashed curve depicts the asymptotic form in the second line in Eqs. \eqref{Sshortt}, while the red dotted curve depicts the large-$t$ asymptotic curve in the second line in Eqs. \eqref{Slarget}, in which we omit the exponential in $t$ relaxation terms. The horizontal 
(black) dashed line indicates the asymptotic value   $\overline{S}_2 = \overline{S_2(\omega=0,t = \infty)} = 4 R_2 k_\mathrm{B} T_2$.
Panel (c): The Pearson coefficient $\rho_0$ as a function of $t$. Solid (blue) curve presents the full exact expression, while the dotted (red) curve indicates the asymptotic form in Eq.~\eqref{asymp}.
	The system parameters are as in Fig.~\ref{fig2}.}
	\label{figA3}
\end{figure}

Comparing the analytical predictions in Eqs. \eqref{Sshortt} with the full curves depicted in Fig. \ref{figA3}, we infer  
that the latter asymptotic forms are valid only for very short observation times - when $t$ is some fraction of a second. 
In turn, in the large-$t$ limit, we get 
\begin{equation}
\label{Slarget}
\begin{split}
\Sigma_1(t) &= 4 R_1 k_\mathrm{B} T_1 + \frac{\alpha_1 k_\mathrm{B} T_1 + \alpha_2 k_\mathrm{B} T_2}{t} + O\left(e^{- t/\tau_r}\right) \,, \\
\Sigma_2(t) &= 4 R_2 k_\mathrm{B} T_2 + \frac{\beta_1 k_\mathrm{B} T_1 + \beta_2 k_\mathrm{B} T_2}{t} + O\left(e^{- t/\tau_r}\right) \,,
\end{split}
\end{equation}
where the symbol $O\left(\exp(- t/\tau_r)\right)$ signifies that the omitted terms decay exponentially with time with $\tau_r$ being some computable constant, and
\begin{equation}
\begin{split}
\alpha_1 &= - \frac{2 R_1^2 \Big(3\left(C + C_1\right)^2 R_1 + \left(4 C^2 + 3 C_1 C_2 + 3 C \left(C_1 + C_2\right)\right) R_2\Big)}{\left(\left(C + C_1\right) R_1 + \left(C+C_2\right) R_2 \right)} \,, \\
\alpha_2 &= \frac{2 R_2 R_1^2 C^2}{\left(\left(C + C_1\right) R_1 + \left(C+C_2\right) R_2 \right)} \,,\\
\beta_1 &= \frac{2 R_1 R_2^2 C^2}{\left(\left(C + C_1\right) R_1 + \left(C+C_2\right) R_2 \right)} \,,\\
\beta_2 &= - \frac{2 R_2^2 \Big(3\left(C + C_2\right)^2 R_2 + \left(4 C^2 + 3 C_1 C_2 + 3 C \left(C_1 + C_2\right)\right) R_1\Big)}{\left(\left(C + C_1\right) R_1 + \left(C+C_2\right) R_2 \right)} \,. 
\end{split}
\end{equation}
It follows from Fig. \ref{figA3} that for the choice of the system's parameter as indicated in the caption, the asymptotic forms in Eqs. \eqref{Slarget} become very accurate already for times $t$ of order $10^{-2}$ ${\rm sec}$. 
We note next that in the general case when $C_1 \neq C_2$ and $R_1 \neq R_2$, the observable 
\begin{align}
\label{parchi}
\overline{\chi}_t = \frac{\Sigma_1(t) }{k_\mathrm{B} R_1} - \frac{\Sigma_2(t)}{k_\mathrm{B} R_2} \sim O\left(\frac{1}{t}\right) \,,
\end{align}
and thus vanishes only in the limit $t \to \infty$.
This substantiates our remark in the text after Eq. \eqref{Rparameter} that some care has to be exercised with the analysis of the limiting behavior and the observation time $t$ has to be taken sufficiently large (of order of a second, as follows from Fig. \ref{figA3}). In the symmetric case with $R_1 = R_2 = R$ and $C_1 = C_2 = \tilde{C}$, the parameter $\chi_t$ is given for any $t$ by
\begin{equation}
\label{arbchi}
\begin{split}
\overline{\chi}_t &= 4  (T_1 - T_2) \Bigg( 1 - \frac{\left(4 C^2 + 6 C \tilde{C} + 3 \tilde{C}^2\right)}{2 \left(C + \tilde{C}\right)} \frac{R}{t}  - \frac{\tilde{C} \left(2 C + \tilde{C}\right) R}{2 \left(C + \tilde{C}\right) t} \exp\left(- \frac{2 \left(C + \tilde{C}\right) t}{\tilde{C} \left(2 C + \tilde{C}\right) R}\right) \\
& + \frac{2 R}{t} \exp\left(- \frac{\left(C + \tilde{C}\right) t}{\tilde{C} \left(2 C + \tilde{C}\right) R}\right)
\Bigg(\left(C+\tilde{C}\right) \cosh\left(\frac{C t}{\tilde{C} \left(2 C + \tilde{C}\right) R}\right) + C \sinh\left(\frac{C t}{\tilde{C} \left(2 C + \tilde{C}\right) R}\right) \Bigg) \Bigg) \,.
\end{split}
\end{equation}
In this case, $\overline{\chi}_t$ is proportional to the difference of the temperatures of the components and thus is exactly equal to zero at any $t$ in equilibrium, i.e., for $T_1 = T_2$, while in the non-equilibrium it approaches a non-zero value.

Lastly, we note that the Pearson coefficient 
is a constant at $t =0$, 
\begin{align}
\rho_0 = \frac{C^2 \left(\left(C + C_2\right) R_2 T_1 + \left(C+C_1\right) R_1 T_2\right)^2}{\left(\left(C+C_2\right)^2 R_2 T_1 + C^2 R_1 T_2\right) \left(\left(C+C_1\right)^2 R_1 T_2 + C^2 R_2 T_1\right)} + O\left(t^2\right) \,,
\end{align}
while its large-$t$ behavior is given by (for both $T_1$ and $T_2$ bounded away from zero)
\begin{align}
\label{rho0as}
\rho_0 = \frac{R_1 R_2 C^2 \Bigg(\Big(\left(C + C_1\right) R_1 + 2 \left(C + C_2\right) R_2 \Big) T_1 + \Big( \left(C + C_2\right) R_2 + 2 \left(C + C_1\right) R_1 \Big) T_2 \Bigg)^2}{4 T_1 T_2 \Big(\left(C + C_1\right) R_1 + \left(C + C_2\right) R_2\Big)^2} \frac{1}{t^2} + O\left(\frac{1}{t^4}\right) \,.
\end{align}
The large-$t$ asymptotic form and the full exact expression for $\rho_0$ are presented in panel (c) of Fig. \ref{figA3}. We observe that for the case at hand the asymptotic form agrees well with the full expression for $t \approx 4 \times 10^{-2}$ ${\rm sec}$.

 We turn finally to the analysis of the exact form of the bivariate probability density function of the spectral densities in case $\omega = 0$. 
Verifying that $|\rho_0| \leq 1$ for any $t$, we are entitled to formally expand the expression in Eq. \eqref{Phif0} in powers of $\rho_0$, which yields the following factorized (with respect to $\lambda_1$ and $\lambda_2$) series
\begin{equation}
\Phi(\lambda_1,\lambda_2) = \frac{1}{\sqrt{\pi}} \sum_{n=0}^{\infty} \frac{\Gamma(n+1/2)}{n!} \rho_0^n \frac{(2 \Sigma_1(t) \lambda_1)^n}{\left(1 + 2 \Sigma_1(t) \lambda_1\right)^{n+1/2}} \frac{(2 \Sigma_2(t) \lambda_2)^n}{\left(1 + 2 \Sigma_2(t) \lambda_2\right)^{n+1/2}} \,.
\end{equation}
Inverting the latter expression we obtain
\begin{equation}
\label{MM}
\begin{split}
P(S_1(\omega=0,t) = s_1, &\, S_2(\omega=0,t) = s_2) = \frac{1}{2 \sqrt{\pi\, \Sigma_1(t) \,\Sigma_2(t)\, s_1 s_2}} \exp\left(- \dfrac{s_1}{2 \Sigma_1(t)} - \dfrac{s_2}{2 \Sigma_2(t)}\right) \\
& \times \sum_{n=0}^{\infty} \frac{n!}{\Gamma(n+1/2)}  L_n^{(-1/2)}\left(\dfrac{s_1}{2 \Sigma_1(t)}\right) L_n^{(-1/2)}\left(\dfrac{s_2}{2 \Sigma_2(t)}\right) \rho_0^n \,.
\end{split}
\end{equation}
This sum is well known and can be performed exactly, which yields our result in Eq. \eqref{dist_voltages}.

\subsection{Probability density of $\chi_t$, defined in Eq.~\eqref{chi}}

Here we seek the probability density function $P(\chi_t)$ of the observable $\chi_t$, defined in~Eq. \eqref{chi}.
To this end,  we first define its moment-generating function: 
\begin{align}
\label{mgfchi}
\Phi_{\chi}(w) = \overline{\exp\left(i w \chi_t\right)} = \overline{\exp\left(i w \left(\frac{S_1(\omega=0,t)}{k_\mathrm{B} R_1} - \frac{S_2(\omega=0,t)}{k_\mathrm{B} R_2}\right)\right)}\,.
\end{align}
The averaging in Eq. \eqref{mgfchi} is most conveniently performed using the factorized series representation 
of the bivariate probability density function $P(S_1(\omega=0,t),S_2(\omega=0,t))$, Eq. \eqref{MM}. Performing some rather straightforward calculations, we find
\begin{align}
\label{chit}
\Phi_{\chi}(w) = \left(1 - 2 i \left(\frac{\Sigma_1(t)}{k_\mathrm{B} R_1} - \frac{\Sigma_2(t)}{k_\mathrm{B} R_2}  \right) w + \frac{4 \left(1 - \rho_0 \right)}{k_\mathrm{B}^2} \frac{\Sigma_1(t)}{R_1}  \frac{\Sigma_2(t)}{R_2} w^2 \right)^{-1/2}
\end{align}
Differentiating the expression \eqref{chit} once and twice, and setting $w=0$, we thus find that the mean obeys Eq. \eqref{parchi}, 
while the variance of the random variable $\chi_t$ is given by
\begin{align}
{\rm Var}\left(\chi_t\right) &= \overline{\chi^2_t} - \overline{\chi_t}^2 = 2 \overline{\chi}_t^2 + \tilde{\chi}^2 \,, \,\,\, \tilde{\chi}^2 = \frac{4 \left(1 - \rho_0 \right)}{k_\mathrm{B}^2} \frac{\Sigma_1(t)}{R_1}  \frac{\Sigma_2(t)}{R_2} \,.
\end{align} 
In turn, the probability density function $P(\chi_t)$ can be evaluated directly from Eq. \eqref{chit} to give
\begin{align}
\label{tilde}
P(\chi_t) &= \frac{1}{2 \pi} \int^{\infty}_{-\infty} \Phi_{\chi}(w) \exp\left( - i w \chi_t\right) dw \nonumber\\
&= \frac{1}{\pi \tilde{\chi}} \exp\left(\frac{\overline{\chi}_t}{\tilde{\chi}^2} \chi_t \right) K_0\left(\frac{\sqrt{\tilde{\chi}^2+ \overline{\chi}_t^2}}{\tilde{\chi}^2} | \chi_t|\right) \,,
\end{align}
where $K_0(x)$ is  the modified Bessel function  of the second kind. 
In the limit $t \to \infty$ the expression simplifies to give our Eq. \eqref{distchi}.

\section{Effective broadness of distributions}
\label{D}

Lastly, we explain why we do call the probability density functions in Eqs. \eqref{distS} and \eqref{P3} effective broad and how this term is to be interpreted in our case. Indeed, in standard nomenclature the term 
"broad" is usually reserved for the distributions with fat power-law tails which are integrable but do not possess moments starting from some order.  In  our case, both distributions do possess moments of an arbitrary positive order and we quantify their effective broadness using a description based on the following argument:

Suppose that we have performed two independent experiments at fixed physical parameters in which we recorded two realizations of $V_1$, and respectively, evaluated two copies of the single-component power spectral density - $S_1^{(1)}(\omega,t)$ and $S_1^{(2)}(\omega,t)$. Then, we define a random variable 
\begin{align}
\label{omega}
W = \frac{S_1^{(1)}(\omega,t)}{S_1^{(1)}(\omega,t)+ S_1^{(2)}(\omega,t)} \,,
\end{align}
which shows us what is the relative contribution of $S_1^{(1)}(\omega,t)$ into the sum of spectral densities of two realizations of the voltages. Intuitively, one may expect that the distribution $P(W)$ of this random variable is peaked at $W = 1/2$, meaning that most probably $S_1^{(1)}(\omega,t)$ and $S_1^{(2)}(\omega,t)$ have the same magnitude. Following Refs. \cite{carlos1,carlos2}, which put forth the concept of such a  random variable in Eq. \eqref{omega} within the context of first-passage time distributions in bounded domains, we call as effectively "narrow" the distributions which are indeed peaked at $W = 1/2$, and as effectively "broad"  - the ones for which it is not the case.

Formally, 
\begin{align}
P(W) = \overline{\delta\left(W - \frac{S_1^{(1)}(\omega,t)}{S_1^{(1)}(\omega,t)+ S_1^{(2)}(\omega,t)}\right)}
\end{align}
Some straightforward calculations (see Refs.~\cite{carlos1,carlos2}) give the following exact expression for $P(W)$:
\begin{align}
P(W) = \frac{1}{(1 - W)^2} \int^{\infty}_0 s_1 ds_1 P(s_1) P\left(\frac{W}{1 - W} s_1\right) \,,
\end{align} 
where $P(s_1)$ is the probability density function of the random variable $S$.  

For $P(s_1)$  in Eq. \eqref{P3} (i.e., for the probability density function of the squared area under $V_1$ divided by $t$), we get
\begin{align}
P(W) = \frac{1}{\pi \sqrt{W (1 - W)}} \,. 
\end{align}
Surprisingly enough, this is the probability density function of the celebrate arcsine law - it diverges when $W \to 0 $ and when $W \to 1$ and has a \textit{minimum} at $W = 1/2$ meaning that an event when $S_1^{(1)}(\omega,t) = S_1^{(2)}(\omega,t)$ is the least probable event and that most likely these random variables have disproportionally different  values: either $S_1^{(1)}(\omega,t) \gg S_1^{(2)}(\omega,t)$ or $S_1^{(1)}(\omega,t) \ll S_1^{(2)}(\omega,t)$.

For $P(s_1)$  in Eq.~\eqref{distS} (i.e., for $S_1(\omega,t=\infty)$) we find
\begin{align}
P(W) \equiv 1,
\end{align}
implying that $W$ is uniformly distributed over its support, so that any relation between $S_1^{(1)}(\omega,t)$ and $S_1^{(2)}(\omega,t)$ is equally probable.

\end{widetext}

\end{document}